\documentclass[nofootinbib,aps,11pt]{revtex4-1}
\usepackage{graphicx}
\usepackage{subfigure} 
\usepackage{hyperref}
\usepackage{cancel}
\usepackage{amssymb}
\usepackage{textcomp}
\usepackage{amsmath}
\usepackage{bm}
\usepackage{times}
\usepackage{epsfig}
\usepackage{color}
\newcommand{\hc}{{\rm h.c.}}
\newcommand{\GeV}{{\rm GeV}}

\newcommand{\SM}{{\rm SM}}
\newcommand{\eq}{{\rm eq}}
\begin{document}
\title{\LARGE Sterile Neutrino Portal Dark Matter from Semi-Production  }	
\bigskip
\author{Ang Liu$^1$}
\author{Feng-Lan Shao$^1$}
\email{shaofl@mail.sdu.edu.cn}
\author{Zhi-Long Han$^2$}
\email{sps\_hanzl@ujn.edu.cn}
\author{Yi Jin$^{2,3}$}
\author{Honglei Li$^2$}
\affiliation{$^1$School of Physics and Physical Engineering, Qufu Normal University, Qufu, Shandong 273165, China\\
	$^2$School of Physics and Technology, University of Jinan, Jinan, Shandong 250022, China
	\\
	$^3$Guangxi Key Laboratory of Nuclear Physics and Nuclear Technology, Guangxi Normal University, Guilin, Guangxi 541004, China}
\date{\today}
\begin{abstract}
In this paper, we study the feeble sterile neutrino portal dark matter under the $Z_3$ symmetry. The dark sector consists of one fermion singlet $\chi$ and one scalar singlet $\chi$, which transforms as $\chi\to e^{i2\pi/3}\chi, \phi\to e^{i2\pi/3}\phi$ under the $Z_3$ symmetry. Regarding fermion singlet $\chi$ as the dark matter candidate, the new interaction terms $y_\chi \phi \bar{\chi^c}\chi$ and $\mu\phi^3/2$ could induce various new production channels. For instance, when $m_\phi>2m_\chi$, the pair decay $\phi\to\chi\chi$ could be the dominant channel, rather than the delayed decay $\phi\to\chi\nu$. Another appealing scenario is when the dark sector is initially produced through the scattering process as $NN\to\chi\chi,NN\to\phi\phi,h\nu\to\chi\phi$, then the semi-production processes $N \chi\to\phi\phi, N\phi\to\phi\chi,N\chi\to\chi\chi$ could lead to the exponential growth of dark sector abundances. The phenomenology of sterile neutrino and the cosmological impact of the dark scalar are also considered in the $Z_3$ symmetric model.

\end{abstract}
\maketitle

\section{Introduction}

The standard model (SM) has made great achievements in particle physics since its establishment, including but not limited to its outstanding interpretation of the basic composition of matter and successful prediction of the Higgs particle \cite{ATLAS:2012yve,CMS:2012qbp}. However, there are still some phenomena that can not be explained by SM, e.g., the origin of tiny neutrino masses and the nature of dark matter (DM). The former is established by the discovery of neutrino oscillation \cite{Kajita:2016cak,McDonald:2016ixn}, which implies that neutrino masses are below the eV scale. The latter is indicated by a variety of evidence, such as the galactic rotation curves, galaxy clusters and large-scale structure of cosmology \cite{Bertone:2016nfn}.

A natural idea is seeking a common interpretation of these two problems, which has been researched  extensively \cite{Krauss:2002px,Asaka:2005an,Ma:2006km,Aoki:2008av,Cai:2017jrq}. Traditionally, high scale sterile neutrinos $N$ are introduced to explain the tiny neutrino mass through the type-I seesaw mechanism \cite{Minkowski:1977sc,Mohapatra:1979ia}. If assuming sterile neutrino has keV-scale mass, it can be regarded as a decaying DM candidate \cite {Dodelson:1993je,Drewes:2016upu,Datta:2021elq,Das:2021nqj}. However, the corresponding parameter space is now tightly constrained by X-ray searches \cite{Ng:2019gch}. One pathway to avoid such constraints is imposing additional symmetry to make the sterile neutrino a stable DM \cite{Ma:2006km,Ma:2007gq}. Then the sterile neutrino becomes the mediator of neutrino mass generation \cite{Hirsch:2013ola}. 

Despite the requirement of large Yukawa coupling and leptogenesis \cite{Davidson:2002qv} favoring high scale sterile neutrinos, the naturalness problem suggests that sterile neutrinos should be below $10^7$ GeV \cite{Vissani:1997ys}. On the other hand, phenomenological studies usually assume that sterile neutrinos are below the TeV scale in order to be detected at colliders \cite{Han:2006ip,Abdullahi:2022jlv}. In this paper, we also consider electroweak scale sterile neutrino. Another advantage of the low scale sterile neutrino is mediating the interaction between the dark matter and SM, which provides new annihilation or production channels of DM \cite{Escudero:2016tzx,Escudero:2016ksa,Campos:2017odj,Batell:2017rol,Hall:2019rld,Coito:2022kif}.

Since particle dark matter was proposed, weakly interacting massive particle (WIMP) is the most popular candidate \cite{Feng:2010gw, Roszkowski:2017nbc, Schumann:2019eaa, Arcadi:2017kky}, which is generated through the freeze-out mechanism. Many experiments are devoted to searching for it through direct or indirect ways \cite{XENON:2018voc,XENON:2023sxq, PandaX-II:2017hlx, PandaX-4T:2021bab, Boveia:2018yeb,Adam:2021rrw, AMS:2013fma,Fermi-LAT:2015att}. Unfortunately, there are no concrete particle DM signals that have been found so far. An alternative candidate is the feebly interacting massive particle (FIMP)\cite{Hall:2009bx,Bernal:2017kxu}, which is produced via the freeze-in mechanism. The interaction between FIMP and SM particles is so weak that it cannot reach the thermal equilibrium state. Consequently, it is produced non-thermally by the decay or annihilation of some particles in the early universe.

The feeble sterile neutrino portal DM under the simplest $Z_2$ symmetry has been studied in Refs. \cite{Bandyopadhyay:2020qpn, Cheng:2020gut, Falkowski:2017uya, Liu:2020mxj,Chang:2021ose}. In this work, we attempt to explore the generation of feeble DM via the sterile neutrino portal with the $Z_3$ symmetry. Within the framework of type-I seesaw, the sterile neutrino $N$ can provide masses for SM neutrinos via the Yukawa interaction $y_\nu \overline{L}\tilde{H}N$, and couples to the dark sector. The dark sector contains a fermion singlet $\chi$ and a scalar singlet $\phi$, both of which transform as $\chi\to e^{i2\pi/3}\chi, \phi\to e^{i2\pi/3}\phi$ under the exact $Z_3$ symmetry.  Providing the mass hierarchy of dark particles as $m_\chi\textless m_\phi$, then the dark fermion $\chi$ becomes a DM candidate. The scenario with strong self-interaction dark scalar $\phi$ and DM produced from the delayed decay $\phi\to \chi \nu$ is studied in Ref. \cite{Ghosh:2023ocl}. Different from this previous study,  we assume that the dark scalar $\phi$ is also feeble interacting with SM. Then we perform a comprehensive investigation of freeze-in production of DM for representative scenarios. The WIMP scenario of sterile neutrino portal DM has also been studied in Ref. \cite{Bandyopadhyay:2022tsf,Liu:2023kil}.

Compared with the $Z_2$ symmetry, the new interactions $\mu \phi^3$ and $y_\chi \phi \bar{\chi}^c\chi$ in this $Z_3$ symmetry will lead to new viable parameter space for DM. Recently, the semi-production of FIMP DM has been proposed in Refs. \cite{Bringmann:2021tjr,Hryczuk:2021qtz}, which can lead to the exponential growth of DM abundance. Semi-production of sterile neutrino DM is then discussed in Ref.~\cite{Bringmann:2022aim}. In this paper, we will show that the exponential growth  of DM via semi-production processes as $N\chi\to\chi\chi$, $N\chi\to\phi\phi$ and $N\phi\to\phi\chi$ is also possible in the $Z_3$ symmetric model.

The structure of this paper is organized as follows. In Sec. \ref{SEC:TM}, we briefly introduce the sterile neutrino portal DM model with the $Z_3$ symmetry. The evolution of feeble DM relic density for some representative scenarios are described in Sec. \ref{SEC:RD}. Then we analyze the constraints from testable signatures under certain scenarios in Sec. \ref{SEC:CP}. Finally, discussions and conclusions are presented in Sec. \ref{SEC:DC}.

\section{The Model}\label{SEC:TM}
\begin{table}
	\begin{center}\Large
		\begin{tabular}{|c| c c c | c c |} 
			\hline
			&  $L$  & ~$N$~ & ~$\chi$~ & ~$H$~ &  ~$\phi$~ \\ \hline
			$SU(2)_L$ & 2 & 1 & 1 & 2 &  1\\ \hline
			$U(1)_Y$  & $-\frac{1}{2}$~  & 0 & 0 & $\frac{1}{2}$ & 0  \\ \hline
			$Z_3$     & 1 & 1 & $\omega$ & 1 & $\omega$ \\
			\hline
		\end{tabular}
	\end{center}
	\caption{Relevant particle contents and the corresponding charge assignments under the  $Z_3$ symmetry. Here $\omega \equiv e^{i 2\pi/3}$.
		\label{Tab:Particle}}
\end{table} 

The sterile neutrino portal DM further extends the SM, which includes the sterile neutrino $N$ and a dark sector with a scalar singlet $\phi$ and a Dirac fermion singlet $\chi$. Among them, $\chi$ is assumed to be the FIMP DM candidate for illustration. The particle contents and the corresponding charge assignments are listed in Table \ref{Tab:Particle}. The exact $Z_3$ symmetry is employed to ensure the stability of DM $\chi$, under which the dark sector fields $\phi$ and $\chi$ transform non-trivially as $\phi \to e^{i2\pi/3}\phi$ and $\chi \to e^{i2\pi/3}\phi$ respectively. Yet the sterile neutrino $N$ and SM fields transform trivially under the $Z_3$ symmetry.  The scalar potential under the unbroken $Z_3$ symmetry is 
\begin{eqnarray} \label{sp}
	V & = & -\mu_{H}^2 H^\dag H
	+\mu_\phi^{2} \phi^\dag \phi + \lambda_H (H^\dag H)^2 + \lambda_\phi (\phi^\dag \phi)^2 + \lambda_{H\phi} (H^\dag H)(\phi^\dag \phi) + \left(\frac{\mu}{2}\phi^3 + h.c.\right),
\end{eqnarray}
where $H$ is the standard Higgs doublet. For simplicity, all the parameters are taken to be real. To guarantee the unbroken $Z_3$ symmetry, $\lambda_\phi \textgreater 0$ and $\mu_\phi \textgreater 0$ must be satisfied. After the electroweak symmetry breaking, $h$ and $\phi$ can obtain physical masses,
\begin{eqnarray}
m_h^2=-2\mu_{H}^2,	
m_\phi^2=\mu_\phi^2+\frac{\lambda_{H\phi}v^2}{2},
\end{eqnarray}
where $h$ is identical to the 125 GeV SM Higgs boson and $v =246~\GeV$. The scalar potential is bounded below with the conditions \cite{Belanger:2012zr}
\begin{equation}
	\lambda_{H} > 0, \quad \lambda_{\phi} > 0, \quad \lambda_{H\phi} + 2 \sqrt{\lambda_{H} \lambda_{\phi}} > 0.
\end{equation}
Meanwhile, the estimation of the lifetime of the desired stable vacuum derives an upper bound on the trilinear coupling, namely $\mu/m_\phi \textless 2\sqrt{\lambda_\phi}$  \cite{Hektor:2019ote}. In the following calculation,   we take $\mu = m_\phi$ and $\lambda_\phi=1$ to meet the above inequality.  

The singlet sterile neutrino $N$ not only provides mass for SM neutrinos through the type-I seesaw mechanism, but also mediates the interaction between SM and the DM. The new Yukawa interactions and mass terms can be written as
\begin{equation}\label{yuk}
	-\mathcal{L}_Y\supset \left(y_\nu\overline{L} \widetilde{H} N  + y_{N} \phi \bar{\chi} N    +\frac{1}{2} m_{N } \overline{N ^c} N + \hc \right)+ y_\chi \phi \bar{\chi^c}\chi+ m_\chi \bar{\chi} \chi,
\end{equation}
where $\widetilde{H}=i\sigma_2 H^{\ast}$. The tiny neutrino mass is generated via the first item, and can be expressed as
\begin{equation}\label{eq:mv}
	m_\nu = - \frac{v^2}{2} y_\nu~ m_{N}^{-1} y_\nu^T.
\end{equation}
To obtain sub-eV scale light neutrino mass, the Yukawa coupling $y_\nu\lesssim\mathcal{O}(10^{-6})$ is required with electroweak scale sterile neutrino $N$. In the following studies, we fix $y_\nu=10^{-6}$ for the benchmark points. The seesaw induced mixing angle between the active and sterile neutrino is then $\theta=y_\nu v/\sqrt{2} m_N\lesssim\mathcal{O}(10^{-6})$.

\section{Relic Density} \label{SEC:RD}

\begin{figure}
	\centering
		\includegraphics[width=0.9\linewidth]{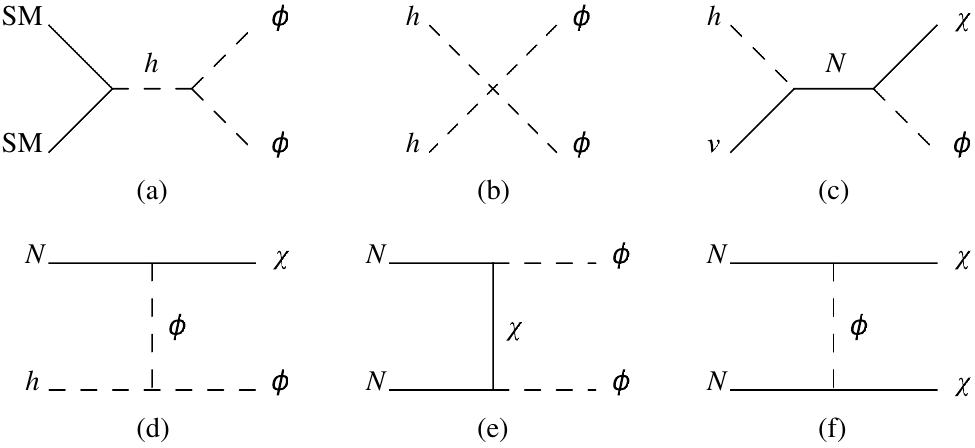}
	\caption{Typical Feynman diagrams for dark sector generation, which also appear in the $Z_2$ symmetric model.}
	\label{FIG:fig1}
\end{figure}

\begin{figure}
	\includegraphics[width=0.9\linewidth]{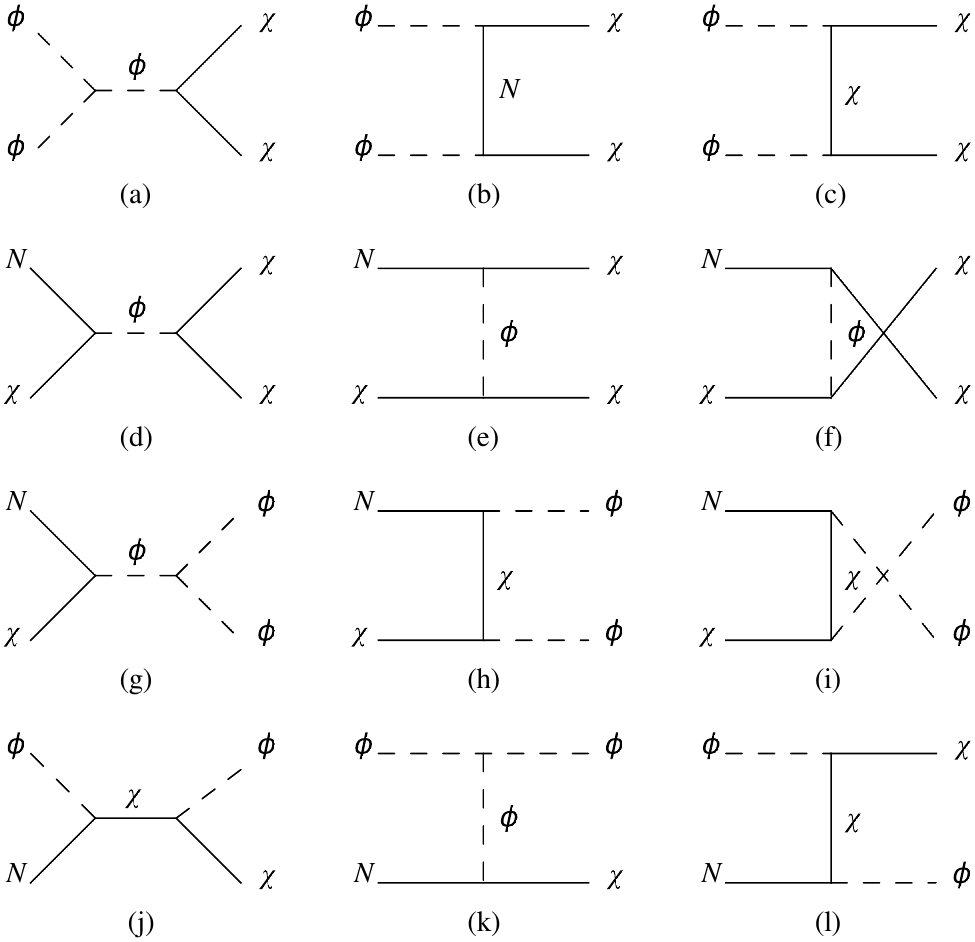}
	\caption{Feynman diagrams for the conversion processes  $\phi\phi\to \chi\chi$, and various semi-production processes $N\chi\to\chi\chi$,  $N\chi\to\phi\phi$, $N\phi\to \phi \chi$.}
	\label{FIG:fig2}
\end{figure}

We consider the fermion singlet $\chi$ as the FIMP DM candidate in this paper. The dark scalar singlet $\phi$ is also assumed  feeble interacting with SM, and is lighter than the sterile neutrino. Meanwhile, the electroweak scale sterile neutrino $N$ is always in thermal equilibrium via neutrino oscillation \cite{Li:2022bpp} or additional interactions \cite{Bandyopadhyay:2022xlp}. The generation of dark scalar $\phi$ is relatively simple, including the Higgs portal annihilation $\SM \to \phi\phi$, the sterile neutrino portal direct decay $N \to \phi \chi$, scattering process $h\nu\to \chi \phi$, $h N\to \chi\phi$, pair annihilation $NN\to\phi\phi$ and semi-production $N\chi \to\phi\phi$.  As for fermion DM $\chi$, it can be produced through plenty of processes, such as direct decay $N \to \phi \chi$, delayed decay $\phi \to \chi \nu$, pair decay $\phi\to \chi\chi$, pair production $N N \to \chi \chi$, semi-production $N \chi \to \chi \chi$, $N\phi\to\phi\chi$, conversion processes $\phi\phi\to\chi\chi$ and so on. In addition to the pair decay $\phi\to \chi\chi$, the semi-production processes $N \chi \to \chi \chi$, $N\chi \to\phi\phi$ and $N\phi \to\phi\chi$ are new in this $Z_3$ symmetric model. Typical Feynman diagrams for dark sector generation and conversion are shown in Figures \ref{FIG:fig1} and \ref{FIG:fig2}. For simplicity, we neglect those channels with petty influences of the relic density of the dark sector, e.g. $h\phi\to\phi\phi$, $h\phi\to\chi\chi$. The relevant Boltzmann equations describing the evolution of dark sector abundances are given by:
\begin{eqnarray}\label{Eqn:BE1}
	\frac{dY_\phi}{dz} &= &  \frac{k}{z^2} \left \langle \sigma v \right \rangle_{h\nu\to\chi\phi}\left(Y_h^{\eq} Y_\nu^{\eq}- \frac{Y_h^{\eq} Y_\nu^{\eq}}{Y_\chi^{\eq} Y_\phi^{\eq}} Y_{\chi} Y_{\phi}\right)
	+\frac{k}{z^2} \left \langle \sigma v \right \rangle_{N h\to\chi\phi}\left(Y_N^{\eq} Y_h^{\eq}- \frac{Y_N^{\eq} Y_h^{\eq}}{Y_\chi^{\eq} Y_\phi^{\eq}} Y_{\chi} Y_{\phi}\right) 
	\nonumber \\
	&+&\frac{k}{z^2} \left \langle \sigma v \right \rangle_{\SM\to\phi\phi}\left(\left(Y_{\SM}^{\eq}\right)^2-\left(\frac{Y_{\SM}^{\eq}}{Y_{\phi}^\eq}\right)^2 Y_{\phi}^2 \right) + \frac{k}{z^2} \left \langle \sigma v \right \rangle_{NN\to\phi\phi}\left(\left(Y_{N}^{\eq}\right)^2-\left(\frac{Y_{N}^{\eq}}{Y_{\phi}^\eq}\right)^2\! Y_{\phi}^2 \right)  \nonumber \\
	&+&k^{\star}z\tilde{\Gamma}_{N\to\phi\chi}\left(Y_{N}^\eq-\frac{Y_{N}^\eq}{Y_{\phi}^\eq Y_{\chi}^\eq}Y_\phi Y_\chi\right) + \frac{k}{z^2} \left \langle \sigma v \right \rangle_{N\chi\to\phi\phi}\left(Y_{N}^{\eq}Y_{\chi}-\frac{Y_{N}^\eq Y_{\chi}^\eq}{(Y_{\phi}^\eq)^2} Y_{\phi}^2\right) \nonumber \\
	&-& \frac{k}{z^2} \left \langle \sigma v \right \rangle_{\phi\phi\to\chi\chi}\left(Y_{\phi}^2-\left(\frac{Y_{\phi}^\eq}{Y_{\chi}^\eq}\right)^2 Y_{\chi}^2\right) - k^{\star}z \tilde{\Gamma}_{\phi\to\chi \nu}\left(Y_{\phi}-\frac{Y_{\phi}^\eq}{Y_{\chi}^\eq}Y_\chi\right) \nonumber \\ 	
	&-& k^{\star}z \tilde{\Gamma}_{\phi\to\chi \chi}\left(Y_{\phi}-\frac{Y_{\phi}^\eq}{(Y_{\chi}^\eq)^2}Y_\chi^2\right)
\end{eqnarray}
\begin{eqnarray}\label{Eqn:BE2}
	\frac{dY_\chi}{dz} &= & 
	  \frac{k}{z^2} \left \langle \sigma v \right \rangle_{h\nu\to\chi\phi}\left(Y_h^{\eq} Y_\nu^{\eq}- \frac{Y_h^{\eq} Y_\nu^{\eq}}{Y_\chi^{\eq} Y_\phi^{\eq}} Y_{\chi} Y_{\phi}\right) 
	+\frac{k}{z^2} \left \langle \sigma v \right \rangle_{N h\to\chi\phi}\left(Y_N^{\eq} Y_h^{\eq}- \frac{Y_N^{\eq} Y_h^{\eq}}{Y_\chi^{\eq} Y_\phi^{\eq}} Y_{\chi} Y_{\phi}\right)
	\nonumber \\
	&+&\frac{k}{z^2} \left \langle \sigma v \right \rangle_{NN\to \chi\chi}\left((Y_{N}^{\eq})^2-\frac{(Y_{N}^\eq)^2}{(Y_{\chi}^\eq)^2} Y_{\chi}^2\right) + \frac{k}{z^2} \left \langle \sigma v \right \rangle_{ N\phi\to \phi\chi}\left(Y_N^{\eq}Y_{\phi}-\frac{Y_N^\eq}{Y_{\chi}^\eq}Y_{\phi} Y_{\chi}\right) 
	\nonumber \\
	&+& \frac{k}{z^2} \left \langle \sigma v \right \rangle_{N\chi\to \chi\chi}\left(Y_{N}^{\eq}Y_{\chi}-\frac{Y_{N}^\eq}{Y_{\chi}^\eq} Y_{\chi}^2\right)
	- \frac{k}{z^2} \left \langle \sigma v \right \rangle_{N\chi\to\phi\phi}\left(Y_{N}^{\eq}Y_{\chi}-\frac{Y_{N}^\eq Y_{\chi}^\eq}{(Y_{\phi}^\eq)^2} Y_{\phi}^2\right)
	\nonumber \\
	&+& \frac{k}{z^2} \left \langle \sigma v \right \rangle_{\phi\phi\to\chi\chi}\left(Y_{\phi}^2-\left(\frac{Y_{\phi}^\eq}{Y_{\chi}^\eq}\right)^2 Y_{\chi}^2\right) + k^{\star}z\tilde{\Gamma}_{N\to\phi\chi}\left(Y_{N}^\eq-\frac{Y_{N}^\eq}{Y_{\phi}^\eq Y_{\chi}^\eq}Y_\phi Y_\chi\right) 
	\nonumber \\
	&+&	k^{\star}z \tilde{\Gamma}_{\phi\to\chi \nu}\left(Y_{\phi}-\frac{Y_{\phi}^\eq}{Y_{\chi}^\eq}Y_\chi\right)+2 k^{\star}z \tilde{\Gamma}_{\phi\to\chi \chi}\left(Y_{\phi}-\frac{Y_{\phi}^\eq}{(Y_{\chi}^\eq)^2}Y_\chi^2\right),
\end{eqnarray}
where we use the definition $z\equiv m_\chi/T$, and $T$ is the temperature. The parameters $k$ and $k^{\star}$ are defined as $k=\sqrt{\pi g_\star/45}m_\chi M_{Pl}$ and $k^{*}=\sqrt{45/4\pi^{3}g_{\star}}M_{Pl}/m_\chi^2$ respectively, where $g_{\star}$ is the effective number of degrees of freedom of the relativistic species and $M_{Pl}=1.2 \times 10^{19}$ GeV is the Planck mass. The thermal decay width $\tilde{\Gamma}_i$ is calculated as $\Gamma_i \mathcal{K}_1/\mathcal{K}_2$ with $\mathcal{K}_{1,2}$ being the first and second modified Bessel Function of the second kind. The corresponding decay widths are given by 
\begin{eqnarray}
	\Gamma_{{N\rightarrow}{\chi\phi}}&=&\frac {y_{N}^{2}}{16\pi{m_N}}{\left(\frac{(m_N+m_\chi)^2-m_\phi^2}{m_N^2}\right)}{\lambda^{1/2}({m_N}^2,{m_\phi}^2,{m_\chi}^2)},  \\
	\Gamma_{\phi\to \chi \nu} &=& \frac {y_{N}^{2}y^2_\nu v^2\, m_\phi}{16\pi m^2_N}{\left(\frac{m_\phi^2-m_\chi^2}{m_\phi^2}\right)^2},  \\
	\Gamma_{\phi\to \chi \chi} &=& \frac {y_{\chi}^{2}}{4\pi{m_\phi^2}}\left(m_\phi^2-4m_\chi^2\right)^{3/2},
\end{eqnarray}
where the kinematic function $\lambda(a,b,c)$ is defined as
\begin{equation} 
	\lambda(a,b,c)=a^2+b^2+c^2-2ab-2ac-2bc.
\end{equation}
Moreover, the thermal average cross sections $\langle \sigma v \rangle$ are calculated numerically by micrOMEGAs \cite{Belanger:2013oya}. For the feeble dark sector, the above Boltzmann equations are solved with the initial condition $Y_\chi=Y_\phi=0$. To avoid possible double counting of generated on-shell particles in the $s$-channel, we also apply the real intermediate states subtraction \cite{Kolb:1979qa}. In the above Boltzmann equations, the dark sector distribution functions following the equilibrium behavior are assumed. More precise calculations involving semi-production processes can be found in Ref.\cite{Bhatia:2023yux}.

The various production channels for DM $\chi$ in this $Z_3$ symmetric model heavily depend on the masses of the dark sector and sterile neutrino. Depending on whether the decays $N\to \phi\chi$ and $\phi\to\chi\chi$ are kinematically allowed, we classify the mass spectrum into four scenarios, namely (1): $m_N>m_\phi+m_\chi$ with $m_\phi<2m_\chi$, (2): $m_N>m_\phi+m_\chi$ with $m_\phi>2m_\chi$, (3): $m_N<m_\phi+m_\chi$ with $m_\phi<2m_\chi$, (4): $m_N<m_\phi+m_\chi$ with $m_\phi>2m_\chi$, where for the latter two scenarios $m_\phi<m_N$ is also satisfied. Theoretically, there are also four scenarios when $m_\phi>m_N$. By replacing the contribution of $N\to \phi\chi$ with $\phi\to N\chi$, we find that the results for $m_\phi>m_N$ scenarios are quite similar to the $m_\phi<m_N$ scenarios, so we will not repeat the $m_\phi>m_N$ scenarios in this paper.

In the following study, we additionally calculate the results under the $Z_2$ symmetry for comparison. Specifically, we give priority to considering benchmark points under the $Z_3$ symmetry to meet the Planck observed relic density $\Omega_\text{DM}h^2=0.12$ \cite{Planck:2018vyg}, whereupon use the parameters occurring under the $Z_2$ symmetry at the same time, i.e. $\{m_\chi, m_\phi, m_N, y_N, y_\nu, \lambda_{H\phi}\}$,  to calculate the abundances of dark particles. In addition, the mass of DM is fixed as 100 GeV for illustration. 

\subsection{Scenario 1}

\begin{table}[h]
	\begin{center}\large
		\begin{tabular}{|c| c| c| c | c |c |c| c| c |} 
			\hline
			Scenario 1 &  $m_\chi$  & ~$m_\phi$~ & ~$m_N$~ & ~$y_N$~ & ~$y_\chi$~ & ~$y_\nu$~  & ~$\lambda_{H\phi}$~ & ~$\mu$~ \\ \hline
			$a$ & 100 & 150 & 300 & $10^{-12}$ & $10^{-12}$ & $10^{-6}$ & $2.1\times10^{-11}$ & 150 \\ \hline
			$b$  & 100  & 150 & 300 & $10^{-12}$  & $5\times10^{-4}$ & $10^{-6}$ & $2.1 \times 10^{-11}$ & 150 \\ \hline
			$c$  & 100  & 150 & 300 & $2.8\times10^{-12}$  & $10^{-12}$ & $10^{-6}$ & $10^{-14}$ & 150 \\ \hline
			$d$  & 100  & 150 & 300 & $2.8\times10^{-12}$  & $5\times10^{-4}$ & $10^{-6}$ & $10^{-14}$ & 150 \\ \hline
		\end{tabular}
	\end{center}
	\caption{The parameter choices for scenario 1, the units of masses involved are GeV.
		\label{Tab:scenario 1}}
\end{table}

\begin{figure}
	\begin{center}
		\includegraphics[width=0.45\linewidth]{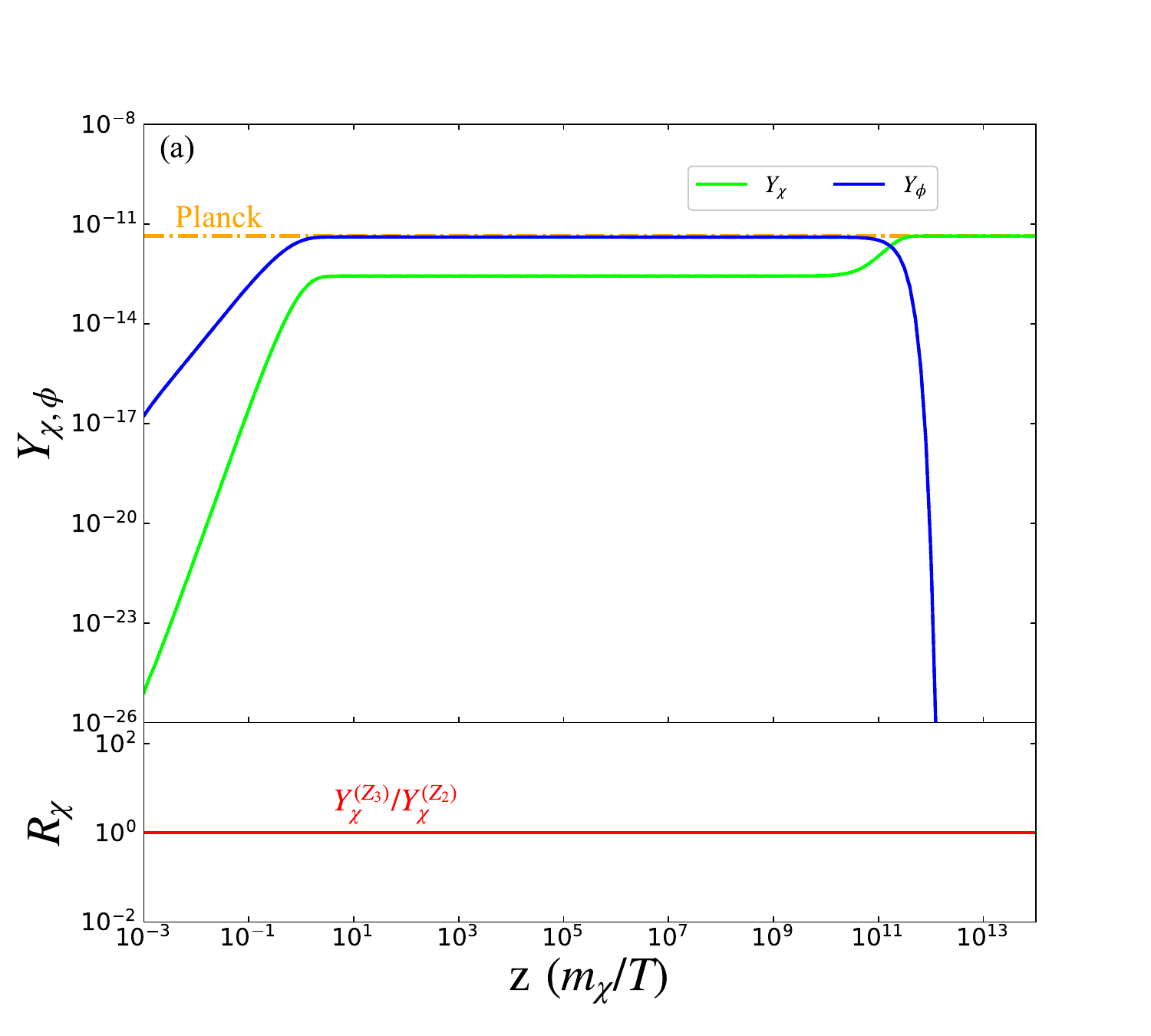}
		\includegraphics[width=0.45\linewidth]{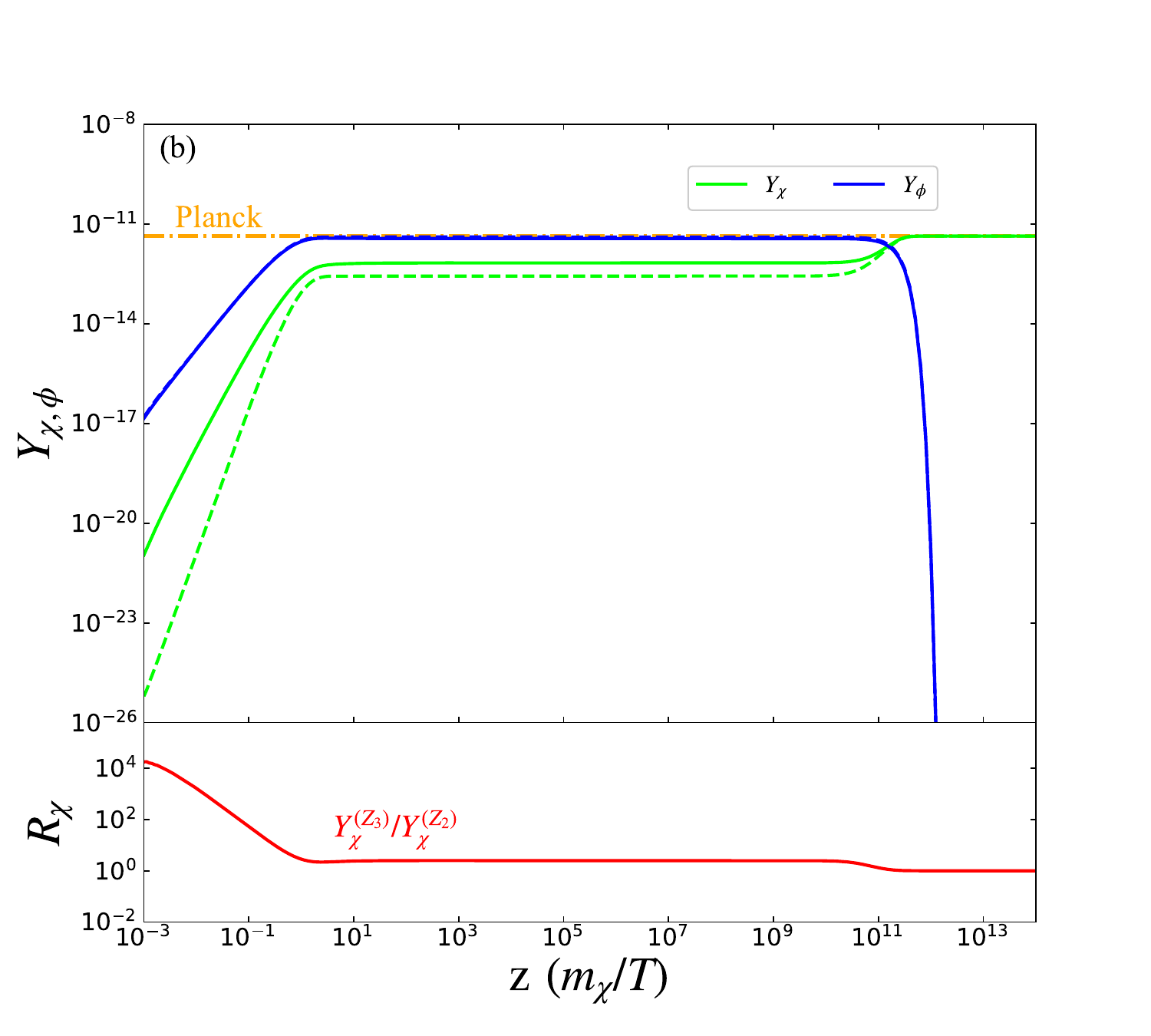}
		\includegraphics[width=0.45\linewidth]{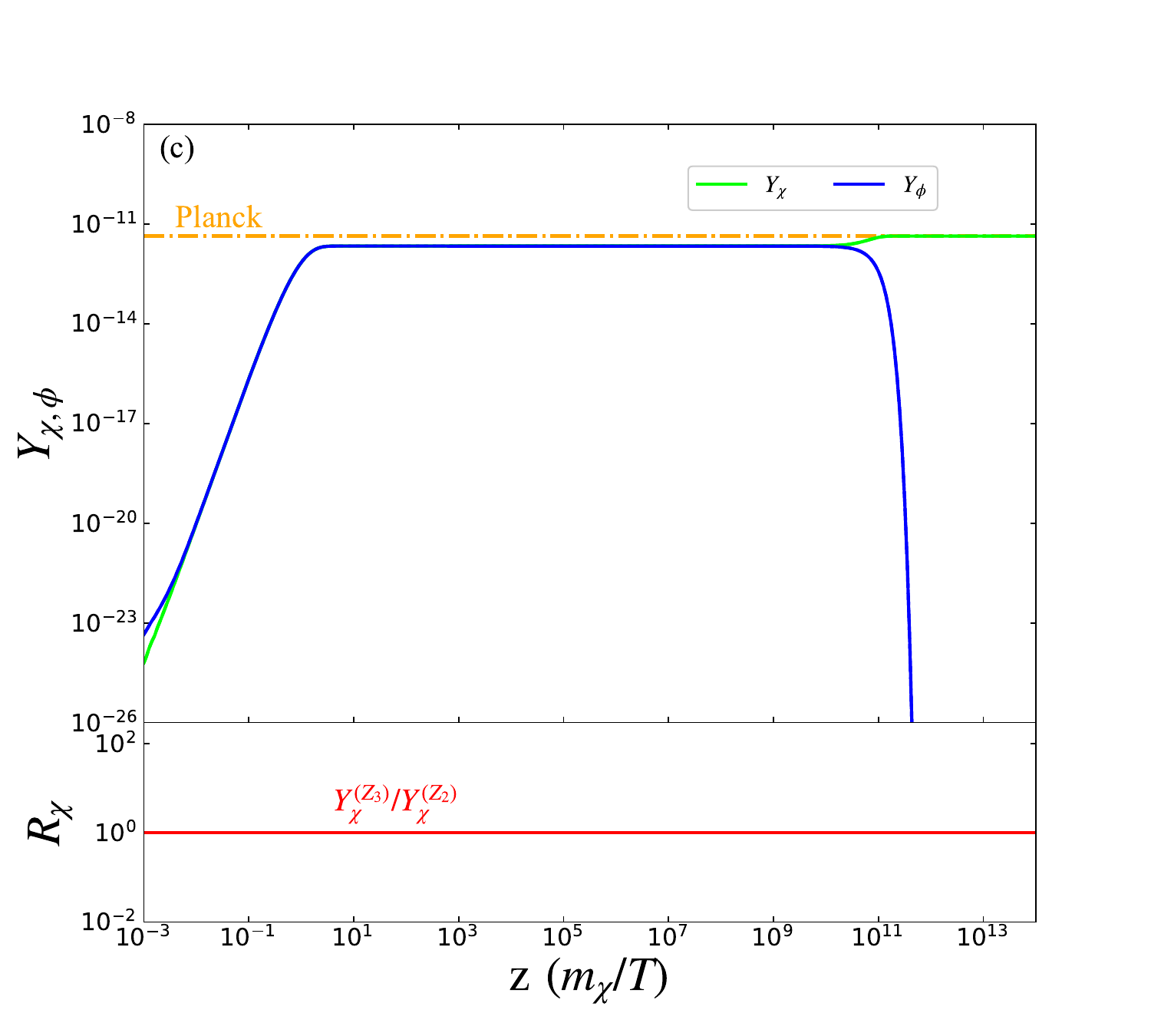}
		\includegraphics[width=0.45\linewidth]{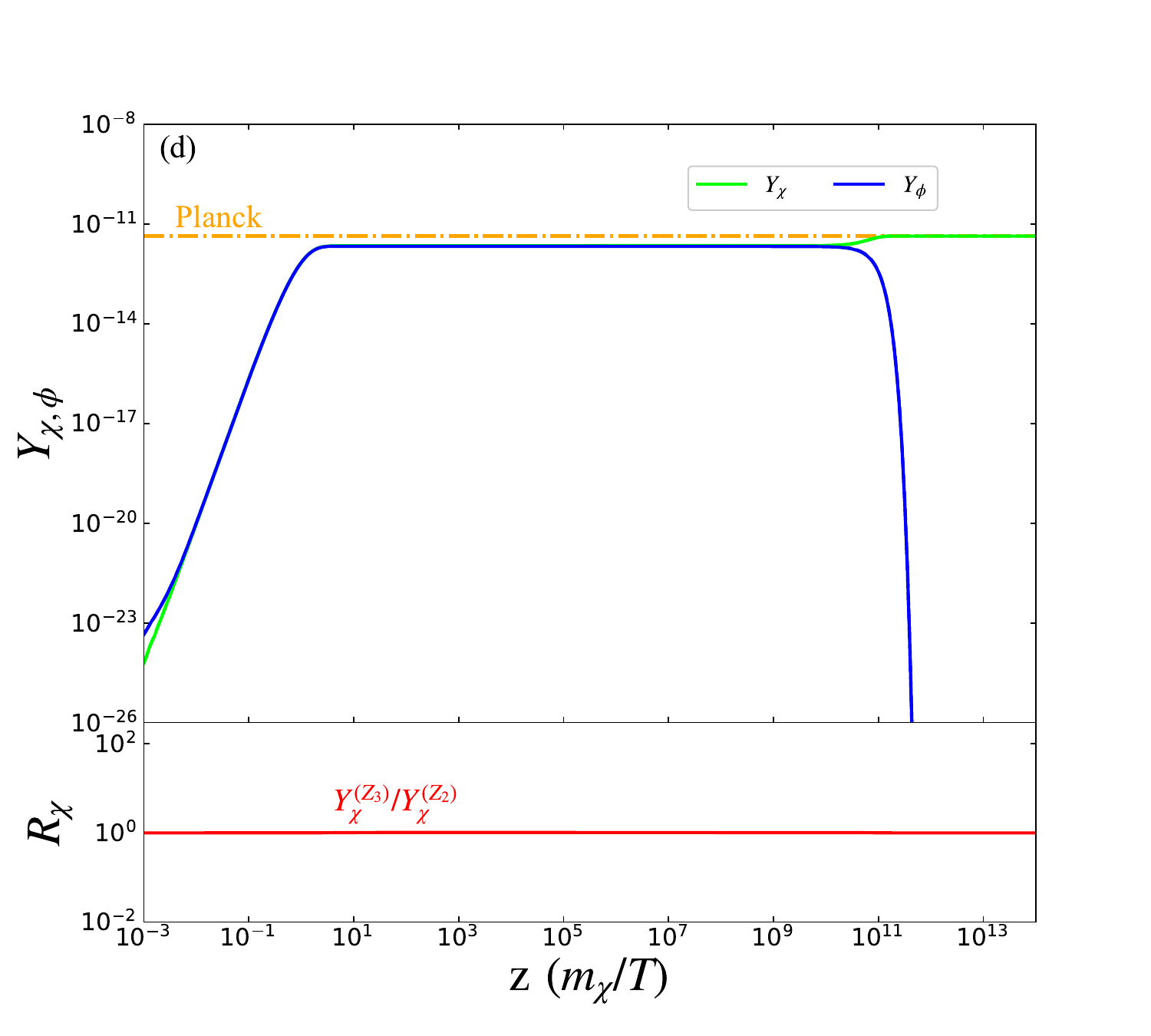}
	\end{center}
	\caption{The evolution of dark sector abundances $Y_\chi$ (green), $Y_\phi$ (blue) and the ratio $R_\chi$ (red) in scenario 1. The solid lines represent the evolution of the dark sector under $Z_3$ symmetry, while the dashed  lines are for the $Z_2$	symmetry. The ratio $R_\chi$ equals $Y_\chi^{(Z_3)}/Y_\chi^{(Z_2)}$, where $Y_\chi^{(Z_3)}$ and $Y_\chi^{(Z_2)}$ are the abundance of DM $\chi$ under the $Z_3$ and the $Z_2$ symmetry respectively. The orange dot-dashed lines are the Planck observed relic density for $m_\chi=100$ GeV. }
	\label{FIG:fig3}
\end{figure}

In scenario 1, we consider that the direct decay $N \to \phi \chi$ is opened, while the pair decay $\phi \to \chi \chi$ is prohibited. The production of dark scalar can be classified into two kinds of process. One is the SM Higgs portal through the coupling $\lambda_{H\phi}$, and the other one is the sterile neutrino portal via the coupling  $y_N$. Meanwhile, the new Yukawa coupling $y_N$ contributes to the conversion processes as shown in Figure~\ref{FIG:fig2}. To illustrate the impact of these conditions, we select four sets of parameters in Table ~\ref{Tab:scenario 1}. The corresponding evolution of $Y_\phi$ and $Y_\chi$ are shown in Figure~\ref{FIG:fig3}.

In scenario 1 (a), we choose the Higgs portal coupling $\lambda_{H\phi}$ being much larger than the sterile neutrino portal coupling $y_N$. In this way, the dark scalar $\phi$ is dominantly generated through the process $\SM\to \phi\phi$, and the decay channel $N \to \phi \chi$ is subdominant. Due to relatively tiny $y_N$ and $y_\chi$, the DM abundance $Y_\chi$ from direct decay $N \to \phi \chi$ is miserly, meanwhile contributions from other $2\to2$ scattering processes are also negligible.  With the cross section $\langle\sigma v\rangle_{\SM\to \phi\phi}\simeq1.9\times10^{-44} ~\rm{cm^{3}/s}$, the Planck observed DM abundance is generated via $\SM\to\phi\phi$ followed by the delayed decay $\phi \to \chi \nu$.
In Figure~\ref{FIG:fig3}~(a), we can see that the evolution of $Y_\chi$ and $Y_\phi$ are consistent in the $Z_2$ and $Z_3$ symmetry all the time,  thus $R_\chi$ equals  one invariably.  This is because of the same generation pattern for the dark sector with only $\phi \to \chi \nu$ allowed in this scenario.

In scenario 1 (b), the value of $y_\chi$ is increased to $5\times10^{-4}$ compared with scenario 1 (a), meanwhile, the other parameters are kept the same. As shown in Figure~\ref{FIG:fig2}, there are new $s$-channel and $t$-channel contributions to the conversion process $\phi\phi\to \chi\chi$ under the $Z_3$ symmetry which do not involve the coupling $y_N$. Different from the $\phi\phi\to\chi\chi$ process, the other conversion processes are suppressed by the smallness of $y_N$. The corresponding cross section  $ \langle \sigma v\rangle_{\phi\phi \to \chi\chi}=6.8\times10^{-29} ~\rm{cm^{3}/s}$ has been greatly enhanced for this scenario, which causes the transition of dark scalar $\phi$ into DM $\chi$. The results are shown in the panel (b) of Figure.~\ref{FIG:fig3}, where $Y_\chi$ is increased by  a factor of 2.5 before $\phi$ decays compared with the $Z_2$ case. According to our calculation, the conversion becomes significant when $y_\chi \gtrsim 10^{-4}$, i.e., the cross section $\langle \sigma v \rangle_{\phi\phi\to\chi\chi} \gtrsim 2.7\times10^{-30} ~\rm{cm^{3}/s}$. The conversion effect leads to the production of DM $\chi$ earlier than the $Z_2$ case. The ratio $R_\chi$ remains on a downward trend until it becomes a constant after $\phi$ totally freeze-in. The value of this constant is proportional to the conversion rate $\langle \sigma v\rangle_{\phi\phi \to \chi\chi}$. In this scenario, the dark scalar $\phi$ is mainly produced via the process $\SM\to \phi\phi$ as in scenario 1 (a), so the same amount of abundance $Y_\phi$ is expected provided the absence of conversion $\phi\phi\to\chi\chi$, which leads to a final reduction of $R_\chi$ to one after the scalar decays via $\phi\to\chi\nu$.

In scenario 1 (c), we consider the opposite case with $\lambda_{H\phi}\ll y_N$. For $\lambda_{H\phi}=10^{-14}$, the Higgs portal process $\SM\to\phi\phi$ is heavily suppressed, so as the other $2\to2$ scattering processes with $y_N\sim y_\chi\sim10^{-12}$. The direct decay $N\to\phi\chi$ becomes the dominant contribution of $Y_\phi$ and $Y_\chi$,  which leads to $Y_\phi=Y_\chi$ at the beginning. The final abundance of dark scalar is then converted into DM via the delayed decay $\phi\to \chi \nu$. In this scenario, the ratio $R_\chi$ equals to one all the time as shown in Figure~\ref{FIG:fig3} (c).

In scenario 1 (d), the conversion process $\phi\phi\to \chi\chi$ is also enhanced with relatively large $y_\chi$. Since the abundance $Y_\phi$ already equals  $Y_\chi$ from $N\to\phi\chi$ as in scenario 1 (c), the strong conversion process does not affect the evolution of the dark sector.

Based on the above results, we can conclude that when the direct decay $N\to \phi\chi$ is allowed and the delayed decay $\phi\to\chi\nu$ is the only decay mode of dark scalar, the final DM abundance in the $Z_3$ symmetric model is the same as in the $Z_2$ symmetric model, although the conversion process $\phi\phi\to\chi\chi$ could impact the evolution of DM. So in scenario 1, we can not distinguish the $Z_3$ symmetry from the $Z_2$ symmetry.

\subsection{Scenario 2}

\begin{table}[h]
	\begin{center}\large
		\begin{tabular}{|c| c| c| c | c |c |c| c| c |} 
			\hline
			Scenario 2 &  $m_\chi$  & ~$m_\phi$~ & ~$m_N$~ & ~$y_N$~ & ~$y_\chi$~ & ~$y_\nu$~  & ~$\lambda_{H\phi}$~ & ~$\mu$~ \\ \hline
			$a$ & 100 & 250 & 400 & $10^{-12}$ & $10^{-12}$ & $10^{-6}$ & $2.0\times10^{-11}$ & 250 \\ \hline
			$b$  & 100  & 250 & 400 & $10^{-12}$  & $5\times10^{-4}$ & $10^{-6}$ & $1.9 \times 10^{-11}$ & 250 \\ \hline
			$c$  & 100  & 250 & 400 & $3.7\times10^{-12}$  & $10^{-12}$ & $10^{-6}$ & $10^{-14}$ & 250 \\ \hline
			$d$  & 100  & 250 & 400 & $3.7\times10^{-12}$  & $5\times10^{-4}$ & $10^{-6}$ & $10^{-14}$ & 250 \\ \hline
		\end{tabular}
	\end{center}
	\caption{The parameter choices for the four cases in scenario 2, the units of masses involved are GeV.
		\label{Tab:scenario 2}}
\end{table}

\begin{figure}
	\begin{center}
		\includegraphics[width=0.45\linewidth]{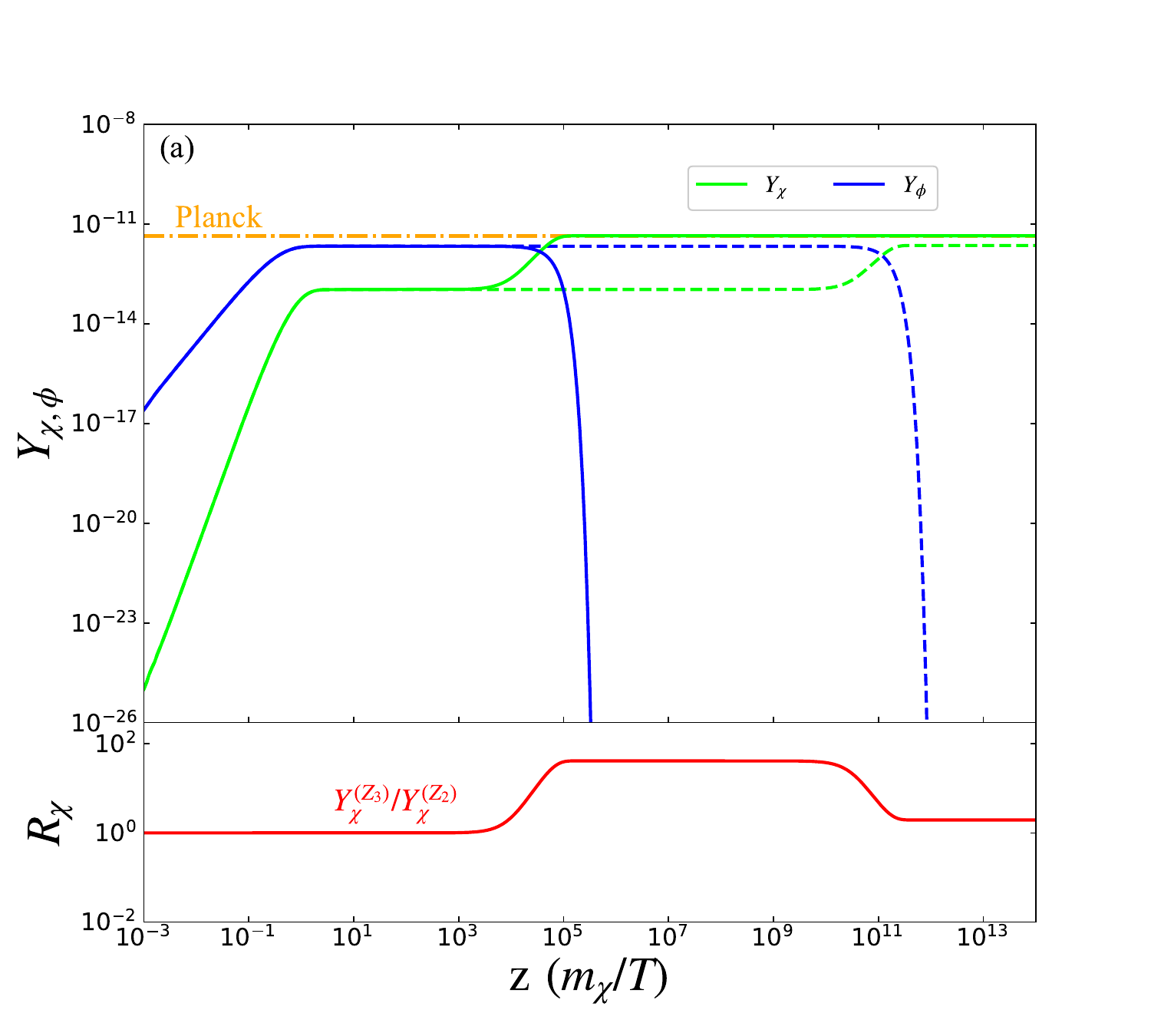}
		\includegraphics[width=0.45\linewidth]{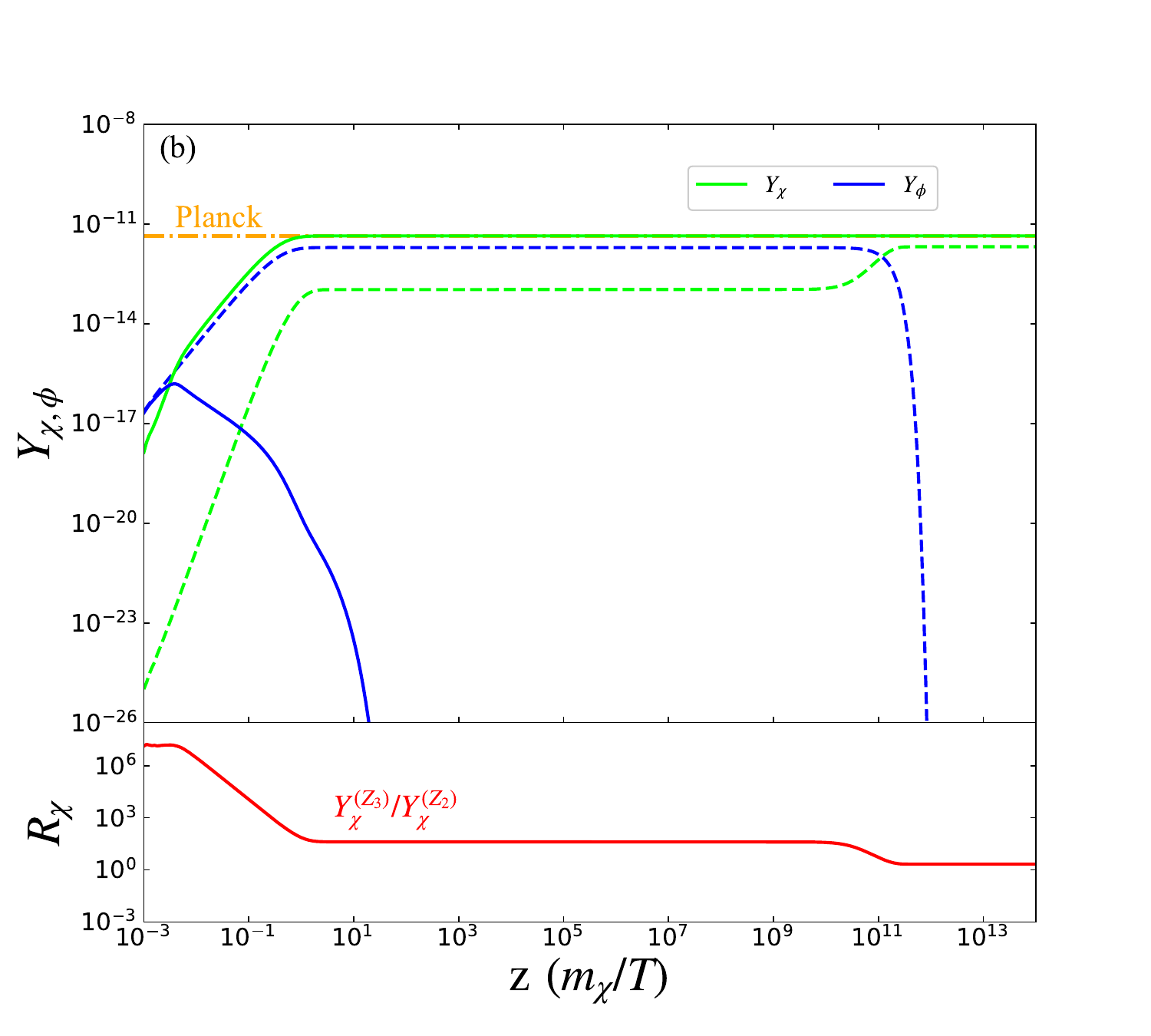}
		\includegraphics[width=0.45\linewidth]{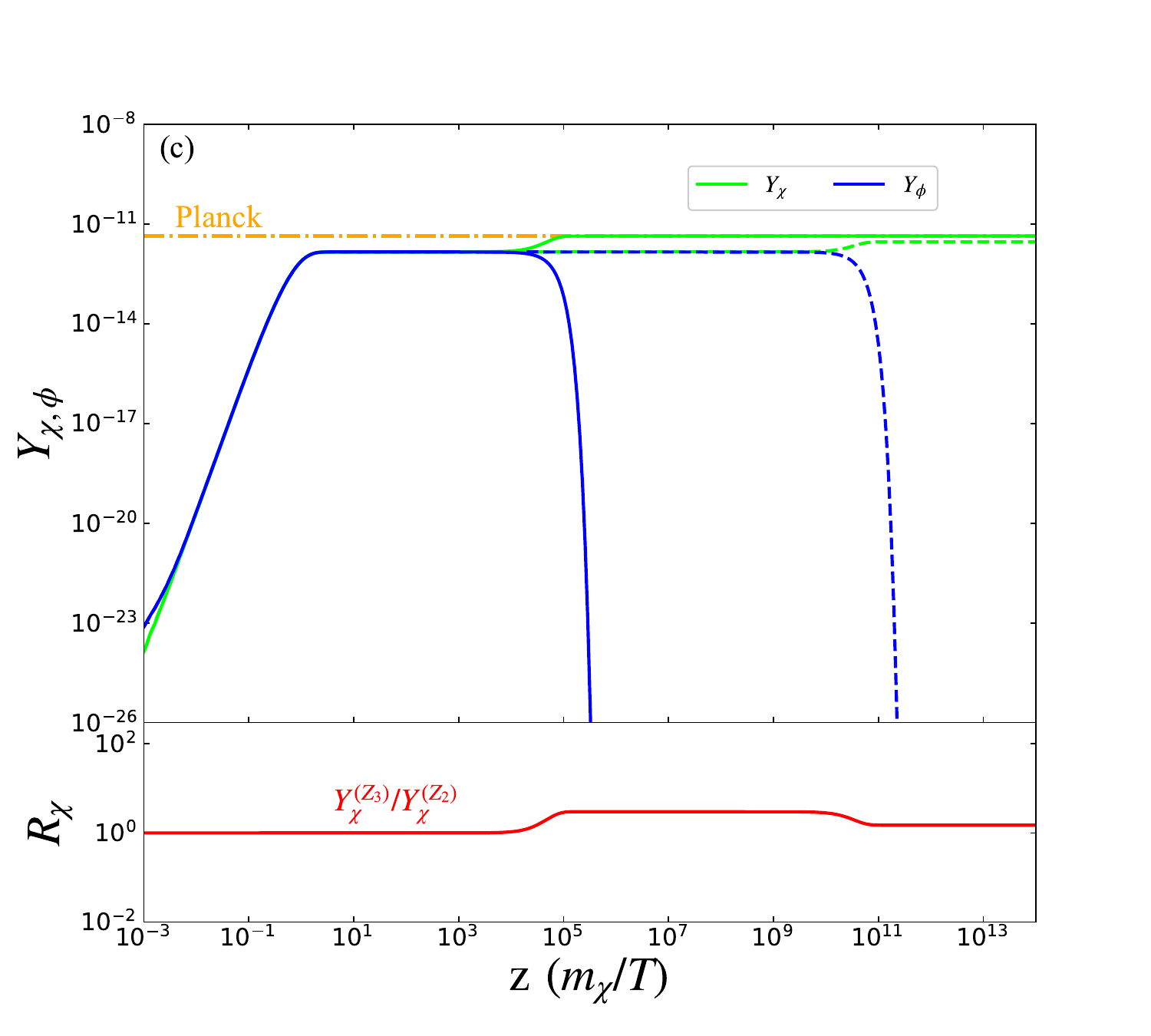}
		\includegraphics[width=0.45\linewidth]{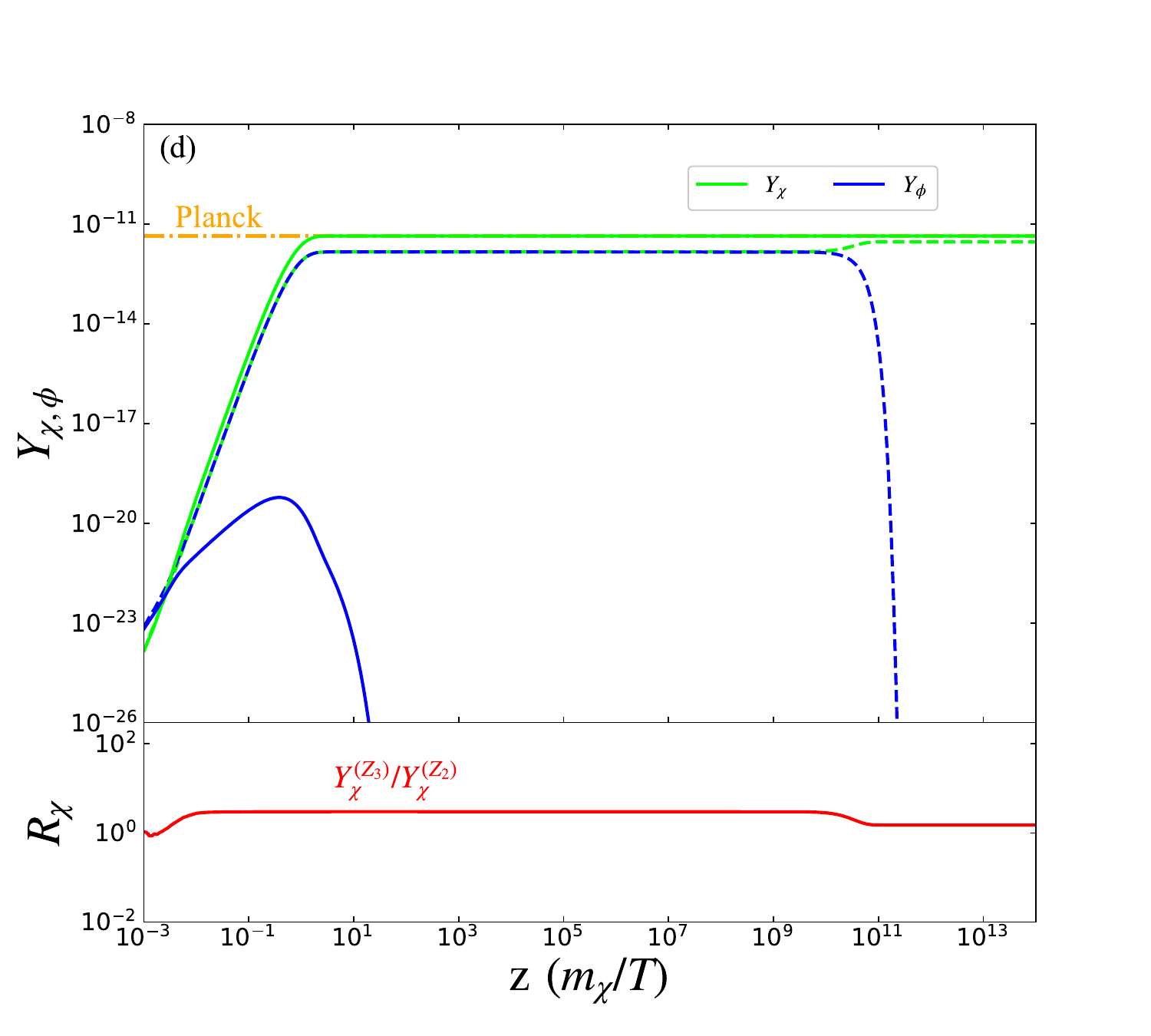}
	\end{center}
	\caption{Same as Figure.~\ref{FIG:fig3}, but for scenario 2.}
	\label{FIG:fig4}
\end{figure}

For scenario 2, we increase the mass of the dark scalar to open the pair decay $\phi \to \chi\chi$, while keeping the decay of $N \to \phi \chi$ allowed. Because the delayed decay $\phi\to\chi\nu$ is further suppressed by the small mixing angle $\theta$, the pair decay $\phi\to\chi\chi$ is the dominant mode even with $y_N\simeq y_\chi$. Four sets of parameters are chosen in Table ~\ref{Tab:scenario 2}. Although the generation mode of dark scalar $\phi$ in scenario 2 is consistent with the corresponding cases in scenario 1, the final conversion of $\phi\to\chi$ is significantly different. Figure.~\ref{FIG:fig4} shows the corresponding evolution of dark particles.

In scenario 2 (a), the contributions from direct decay $N\to\phi\chi$ to the dark sector abundances are tiny. The dark scalar $\phi$ is dominantly produced from $\SM\to \phi\phi$. Correct abundance $Y_\chi$ is obtained with $\langle\sigma v\rangle_{\SM\to \phi\phi}\simeq6.3\times10^{-45} ~\rm{cm^{3}/s}$ followed by the pair decay $\phi\to\chi\chi$. The conversion of $\phi\to\chi$ happens much earlier than the $Z_2$ symmetric model due to $\Gamma_{\phi\to \chi \chi}\gg \Gamma_{\phi\to \chi\nu}$. The ratio $R_\chi$ equals  one  before $\phi$ decays, and quickly increases to 41.6 after $\phi$ decays. Since this pair decay converts one $\phi$ into two $\chi$, the observed DM abundance $Y_\chi^\text{obs}$ is realized with $Y_\phi(z=10)=Y_\chi^\text{obs}/2$ in the $Z_3$ symmetric model. In the $Z_2$ symmetric model, the conversion is via the delayed decay $\phi\to\chi\nu$, which leads to $Y_\chi(z=\infty)=Y_\phi(z=10)=Y_\chi^\text{obs}/2$. So the  final ratio $R_\chi$ is two in scenario 2 (a). 

In scenario 2 (b), the relatively large $y_\chi$ not only enhances the conversion rate of $\phi\phi\to\chi\chi$, but also increases the decay width $\Gamma_{\phi\to \chi \chi}$.  Our numerical calculation finds that compared with scenario 2 (a), a slightly smaller $\lambda_{H\phi}$ with $\langle\sigma v\rangle_{\SM\to \phi\phi}\simeq5.7\times10^{-45} ~\rm{cm^{3}/s}$ could satisfy the Planck constraint. Once produced, the dark scalar decays quite quickly into a DM pair, which results in $Y_\phi\ll Y_\chi$. The inverse conversion process and the fast pair decay transform a small part of the dark sector as $2\chi\to2\phi\stackrel{\rm decay}{\longrightarrow}4\chi$, which makes the generation of DM more efficient in this scenario. The ratio $R_\chi$ decreases during the evolution, and finally  $R_\chi$ reaches about 2.1 in scenario 2 (b).

In scenario 2 (c), the dark sector abundances $Y_\phi$ and $Y_\chi$ are initially produced via the direct decay $N\to \phi\chi$. Then the dark scalar $\phi$ is converted to DM $\chi$ by the pair decay $\phi\to\chi\chi$. The cascade decay chain is $N\to \phi\chi\to\chi\chi\chi$ in the $Z_3$ symmetric model. Under the $Z_2$ symmetry, the decay chain is $N\to \phi\chi\to \chi\nu\chi$. So as shown in Figure \ref{FIG:fig4} (c), the ratio $R_\chi$ increases to 3 after $\phi$ decays in the $Z_3$ symmetric model, and then decreases to 3/2 after $\phi$ decays in the $Z_2$ symmetric model.

In scenario 2 (d), the initial dark sector abundances from $N\to\phi\chi$ decay are much smaller than in scenario 2 (b), so the contribution from the conversion process $\phi\phi\to\chi\chi$ is too small to make $Y_\chi$ exceed obviously even with the same $y_N$. Therefore, the increase of $R_\chi$ in the early stage is mainly determined by $\phi \to \chi\chi$. The final ratio $R_\chi$ is also 3/2 in scenario 2 (d).

The new pair decay $\phi\to\chi\chi$ makes the $Z_3$ symmetric model different from the $Z_2$ symmetric model. With the same couplings in the $Z_3$ symmetric model, the generated DM abundance in the $Z_2$ symmetric model is always smaller than the observed value. Depending on the dominant generation process of dark scalar, the ratio $R_\chi$ is also different. When the dark scalar is dominantly produced via the Higgs portal $\SM\to\phi\phi$, the final ratio is $R_\chi\gtrsim 2$. Meanwhile, if the dark scalar is generated from direct decay $N\to\phi\chi$, the predicted final ratio is $R_\chi=3/2$. The dark scalar is short-lived in the $Z_3$ symmetric model due to the relatively large partial decay width $\Gamma_{\phi\to\chi\chi}$. Then the tight constraints from cosmology can be easily satisfied in scenario 2. 
 
\subsection{Scenario 3}

\begin{table}[h]
	\begin{center}\large
		\begin{tabular}{|c| c| c| c | c |c |c| c| c |} 
			\hline
			Scenario 3 &  $m_\chi$  & ~$m_\phi$~ & ~$m_N$~ & ~$y_N$~ & ~$y_\chi$~ & ~$y_\nu$~  & ~$\lambda_{H\phi}$~ & ~$\mu$~ \\ \hline
			$a$ & 100 & 140 & 200 & $10^{-12}$ & $10^{-12}$ & $10^{-6}$ & $2.2\times10^{-11}$ & 140 \\ \hline
			$b$  & 100  & 140 & 200 & $10^{-12}$  & $5\times10^{-4}$ & $10^{-6}$ & $2.2 \times 10^{-11}$ & 140 \\ \hline
			$c$  & 100  & 140 & 200 & $4.5\times10^{-7}$  & $10^{-12}$ & $10^{-6}$ & $10^{-14}$ & 140 \\ \hline
			$d$  & 100  & 140 & 200 & $1.5\times10^{-7}$  & $6.2\times10^{-1}$ & $10^{-6}$ & $10^{-14}$ & 140 \\ \hline
		\end{tabular}
	\end{center}
	\caption{The parameter choices for the four cases in scenario 3, the units of masses involved are GeV.
		\label{Tab:scenario 3}}
\end{table}

\begin{figure}
	\begin{center}
		\includegraphics[width=0.45\linewidth]{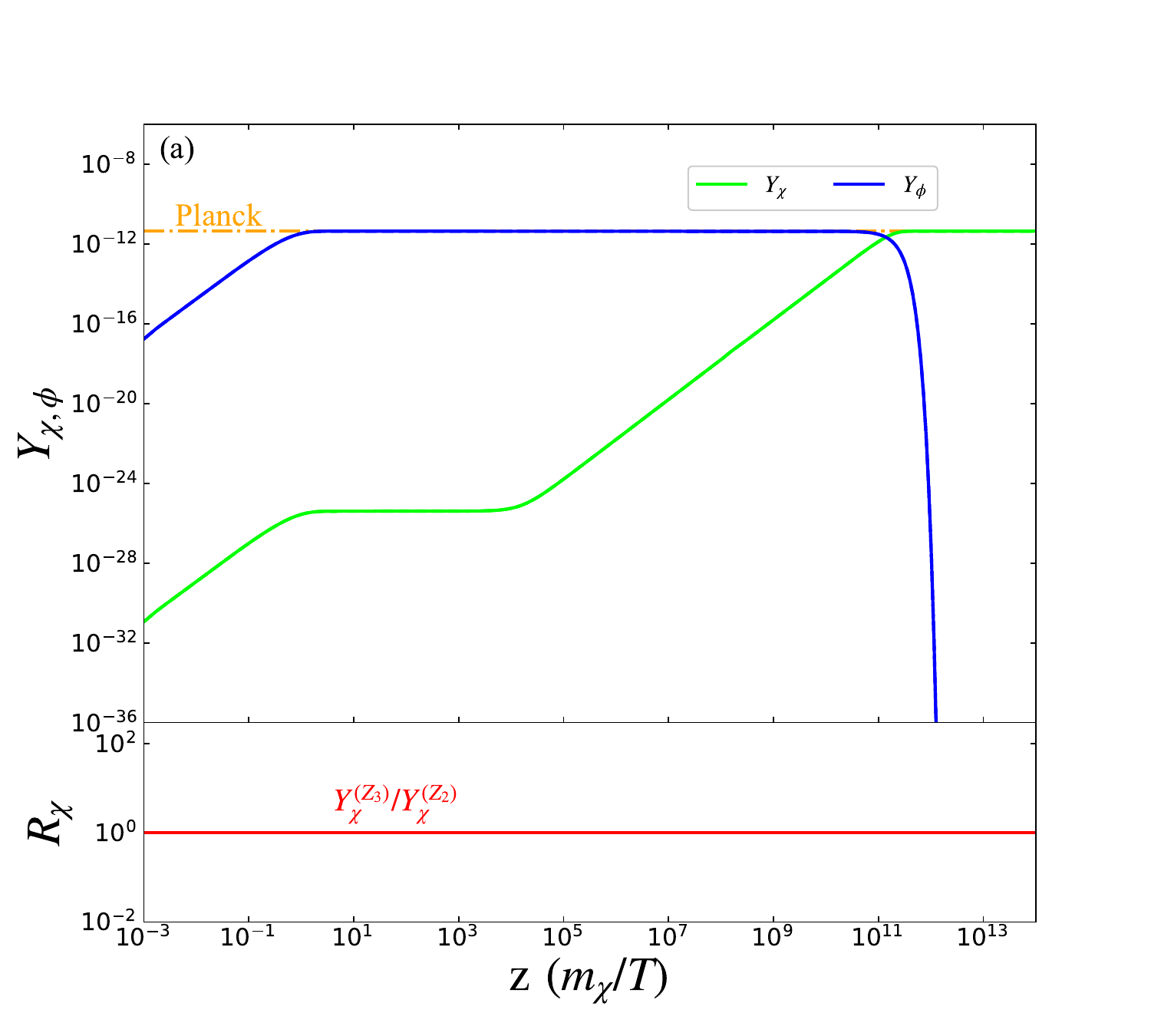}
		\includegraphics[width=0.45\linewidth]{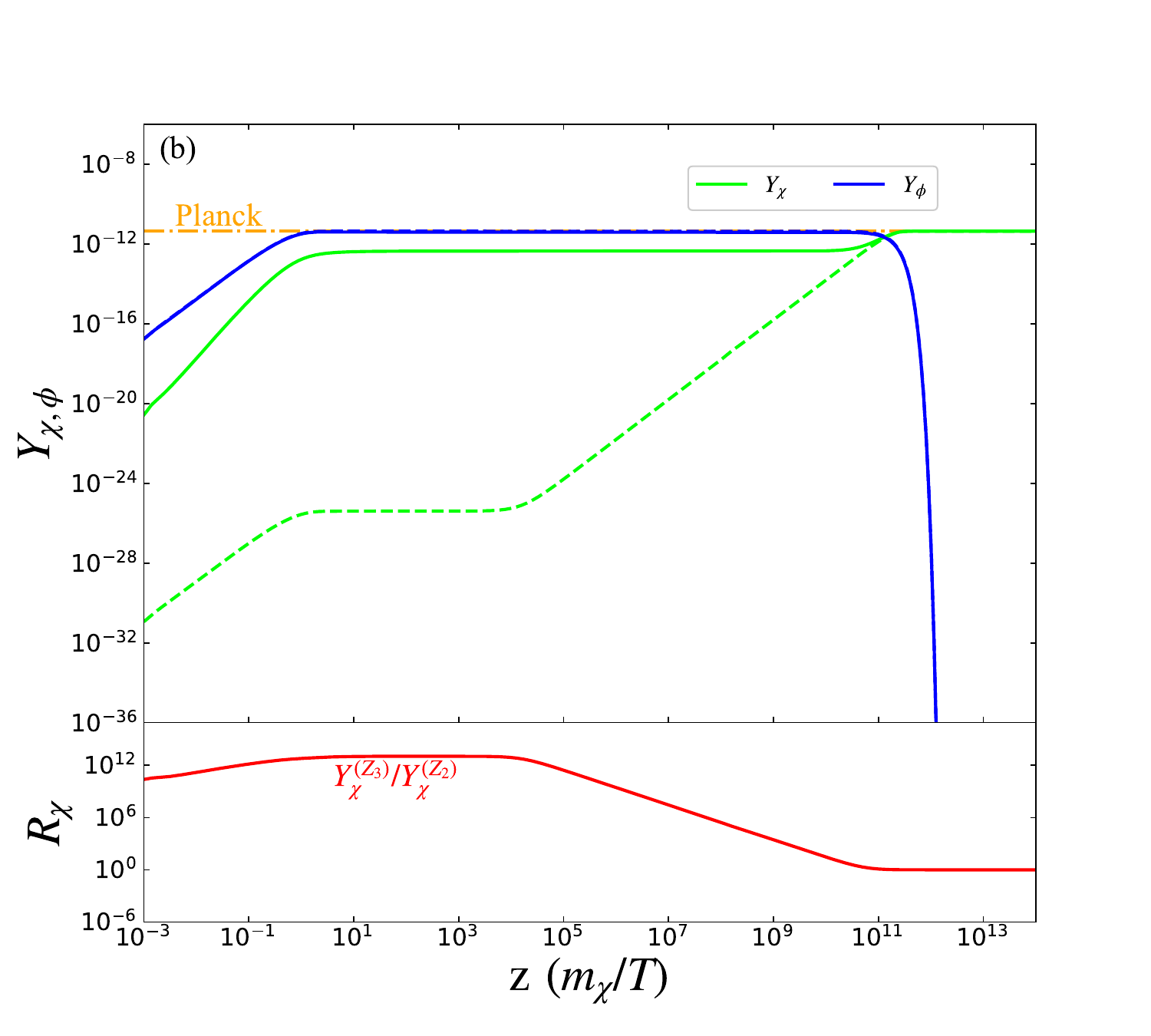}
		\includegraphics[width=0.45\linewidth]{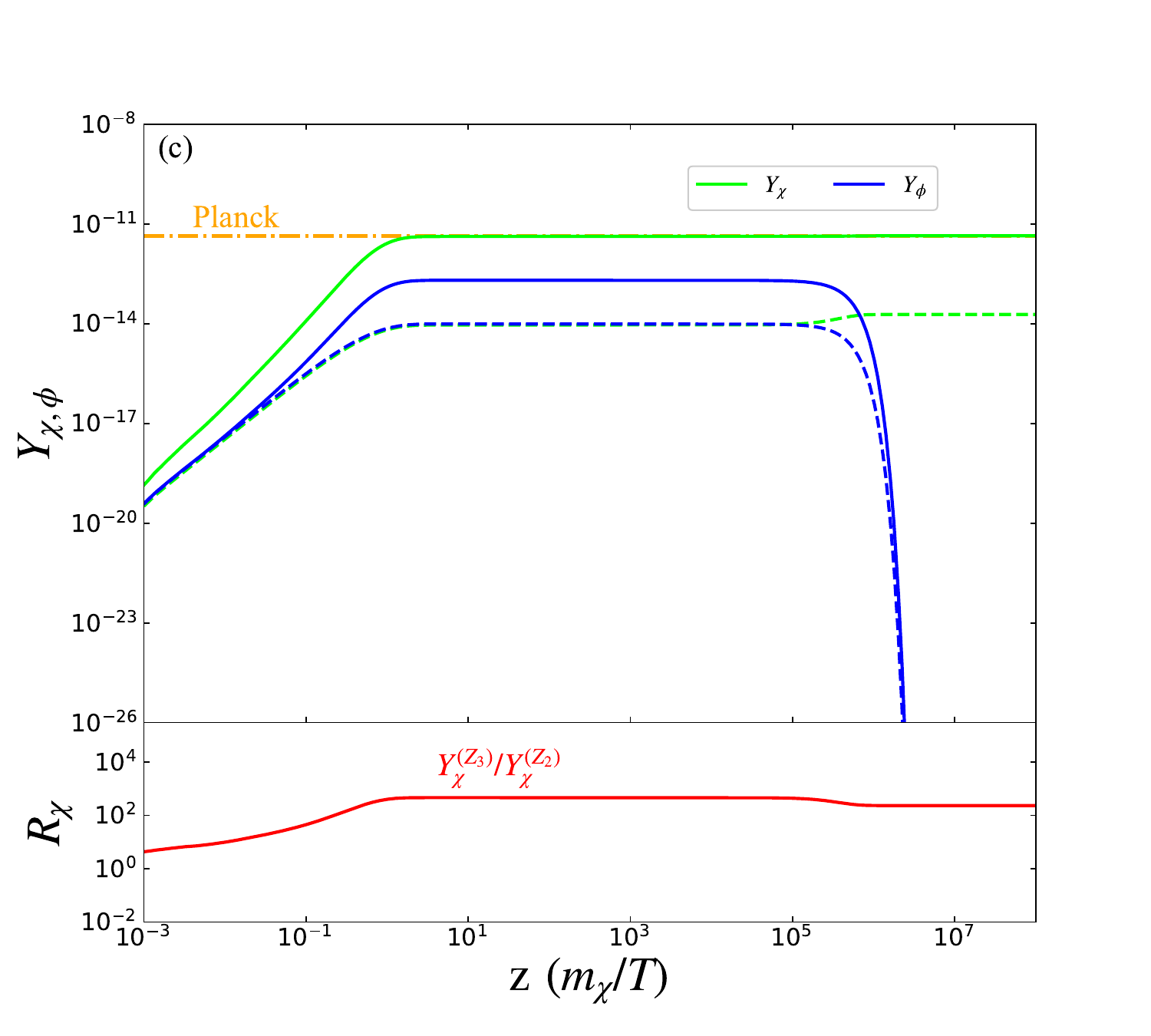}
		\includegraphics[width=0.45\linewidth]{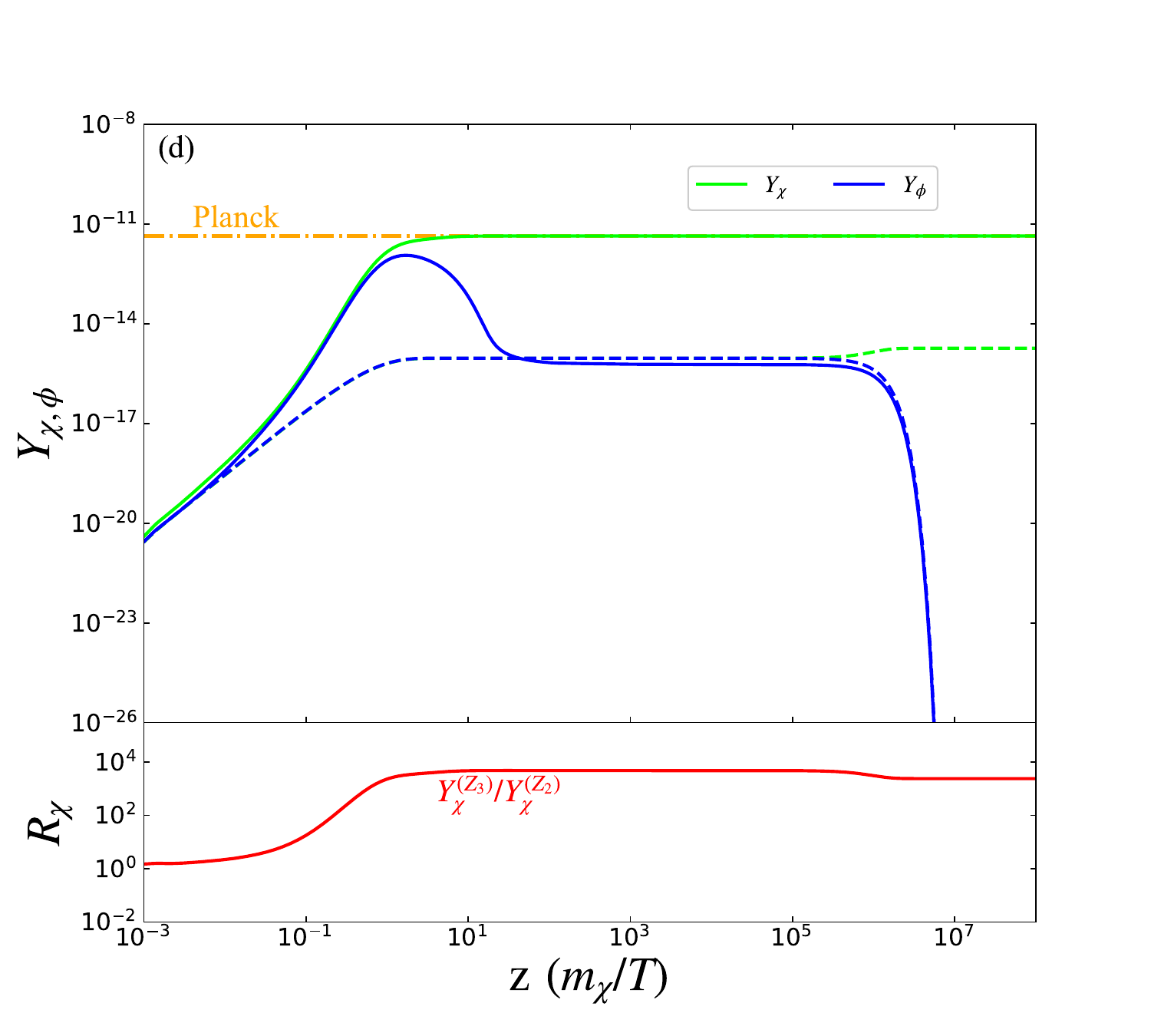}
	\end{center}
	\caption{Same as Figure.~\ref{FIG:fig3}, but for scenario 3.}
	\label{FIG:fig5}
\end{figure}

The sterile neutrino portal coupling $y_N$ is at the order of $\mathcal{O}(10^{-12})$ aiming not to exceed the observed DM relic abundance from direct decay $N\to\phi\chi$ in the previous two scenarios. In scenario 3, we consider that both $N \to \phi\chi$ and $\phi \to \chi\chi$ are prohibited kinematically. Compared to the previous two scenarios, the $2\to2$ scattering channels as $NN\to\chi\chi$ and $h\nu\to \chi\phi$ will dominate the production of $\chi$ at the very beginning in this scenario. Besides the Higgs portal $\SM\to\phi\phi$ channels, the other scattering processes can also make considerable contributions to the production of $\phi$.   We take four sets of parameters in Table ~\ref{Tab:scenario 3} to illustrate this scenario. In addition, the evolution of the abundance of dark particles is shown in Figure~\ref{FIG:fig5}.

In scenario 3 (a),  the dark scalar $\phi$ is dominantly produced via $\SM\to\phi\phi$. With $\langle \sigma v \rangle_{\SM\to\phi\phi}\simeq2.4\times10^{-44} ~\rm{cm^{3}/s}$, correct DM relic abundance $Y_\chi$ is obtained by delayed decay $\phi\to\chi\nu$. It is obvious in Figure \ref{FIG:fig5} (a) that the contribution from scattering to the generation of DM $\chi$ is much lower than that from $N\to\phi\chi$ decay. Without the contribution from direct decay $N\to\phi\chi$ to $Y_\phi$, a slightly larger $\lambda_{H\phi}$ is required compared with scenario 1 (a). The ratio $R_\chi$ is invariant to one due to the same transformation process under the two symmetries.

In scenario 3 (b), the conversion process  $\phi\phi \to \chi\chi$ is enhanced, which becomes the dominant production mode of $\chi$. The large conversion rate leads $R_\chi$ to rise to an enormous value  $\sim\mathcal{O}(10^{13})$ in the initial time, and then decrease to one with the completion of $\phi \to \chi\nu$.

In scenario 3 (c), the contribution of $\SM\to\phi\phi$ can be ignored due to tiny $\lambda_{H\phi}$. The dark sector is primarily generated by scattering processes as $NN \to \chi\chi, NN\to \phi\phi$, $h\nu\to\chi\phi$ at the very beginning. The typical scattering cross sections are $\langle \sigma v \rangle_{NN \to \chi\chi} \simeq1.2\times10^{-48} ~\rm{cm^{3}/s}$, $\langle \sigma v \rangle_{NN \to \phi\phi} \simeq1.8\times10^{-48} ~\rm{cm^{3}/s}$ and $\langle \sigma v \rangle_{\chi\phi\to h\nu} \simeq2.6\times10^{-47} ~\rm{cm^{3}/s}$ for the benchmark point.  It can be seen from Figure.~\ref{FIG:fig5} (c) that the generated dark abundances from scattering are two orders of magnitudes lower than the observed value under the $Z_2$ symmetry. Nevertheless, the new semi-production processes  $N\chi\to\phi\phi$ and $N\phi \to \phi\chi$ are enhanced with $\mu=m_\phi$ and $y_N=4.5\times10^{-7}$ under the $Z_3$ symmetry, which results in the exponential growth of dark sector abundances. It is worth mentioning that the assumption of thermal equilibrium of sterile neutrino is important to realize such exponential growth \cite{Bringmann:2021tjr}. For the benchmark point, the DM abundance $Y_\chi$ is much larger than the dark scalar abundance $Y_\phi$, so the contribution from delayed decay $\phi\to\chi\nu$ to the total $Y_\chi$ is not obvious. Naturally, the ratio $R_\chi$ exponentially increases to $R_\chi^\text{max} \simeq4.6\times10^2$ until the end of the semi-production processes. Afterwards $R_\chi$ is affected by $\phi \to \chi\nu$, and finally decreases to $2.3\times10^2$ in scenario 3 (c).

In scenario 3 (d), we reduce the value of $y_N$, so $Y_\chi$ will eventually fail to satisfy the observed relic density even with the enhancement by the semi-production processes $N\chi\to\phi\phi$ and $N\phi \to \phi\chi$ as in scenario 3 (c). On the other hand, $y_\chi$ is taken as a large value $6.2\times10^{-1}$, which then increases the third semi-production processes $N\chi \to \chi\chi$ with $\langle\sigma v \rangle_{N\chi \to \chi\chi}\simeq2.6\times10^{-36} ~\rm{cm^{3}/s}$. The new semi-production process $N\chi \to \chi\chi$ will cause additional contribution to the exponential growth of $Y_\chi$ to satisfy the Planck constraint. Meanwhile, the cross section of conversion process $\phi\phi \to \chi \chi$ is greatly enhanced to about $1.0\times10^{-22} ~\rm{cm^{3}/s}$, which makes an equal amount of $Y_\phi$ and $Y_\chi$ when $z\lesssim1$. Afterward, the conversion process quickly converts the dark scalar into DM.  The ratio $R_\chi$ exponentially increases to $R_\chi^\text{max} \simeq4.8\times10^3$, and $R_\chi$ finally decreases to $2.4\times10^3$.

The former two cases in scenario 3 indicate that when the DM abundance is dominant by the delayed decay $\phi\to\chi\nu$, the predicted final DM abundances of $Z_2$ and $Z_3$ are the same. However, when DM is primarily generated through the neutrino portal scattering process  $NN\to\chi\chi$ and $h\nu\to\phi\chi$, the semi-production processes $N\chi\to\phi\phi$, $N\phi \to \phi\chi$ and $N\chi \to \chi\chi$ could lead to the exponential growth of the dark sector abundances. The latter two cases in scenario 3 have quite different predictions between the $Z_2$ and $Z_3$ symmetric models, thus are useful to distinguish these two models.

\subsection{Scenario 4}

\begin{table}[h]
	\begin{center}\large
		\begin{tabular}{|c| c| c| c | c |c |c| c| c |} 
			\hline
			Scenario 4 &  $m_\chi$  & ~$m_\phi$~ & ~$m_N$~ & ~$y_N$~ & ~$y_\chi$~ & ~$y_\nu$~  & ~$\lambda_{H\phi}$~ & ~$\mu$~ \\ \hline
			$a$ & 100 & 250 & 300 & $10^{-12}$ & $10^{-12}$ & $10^{-6}$ & $2.1\times10^{-11}$ & 250 \\ \hline
			$b$  & 100  & 250 & 300 & $10^{-12}$  & $5\times10^{-4}$ & $10^{-6}$ & $2.0 \times 10^{-11}$ & 250 \\ \hline
			$c$  & 100  & 250 & 300 & $3.2\times10^{-7}$  & $10^{-12}$ & $10^{-6}$ & $10^{-14}$ & 250 \\ \hline
			$d$  & 100  & 250 & 300 & $4.6\times10^{-7}$  & $5\times10^{-4}$ & $10^{-6}$ & $10^{-14}$ & 250 \\ \hline
		\end{tabular}
	\end{center}
	\caption{The parameter choices for the four cases in scenario 4, the units of masses involved are GeV.
		\label{Tab:scenario 4}}
\end{table}

\begin{figure}
	\begin{center}
		\includegraphics[width=0.45\linewidth]{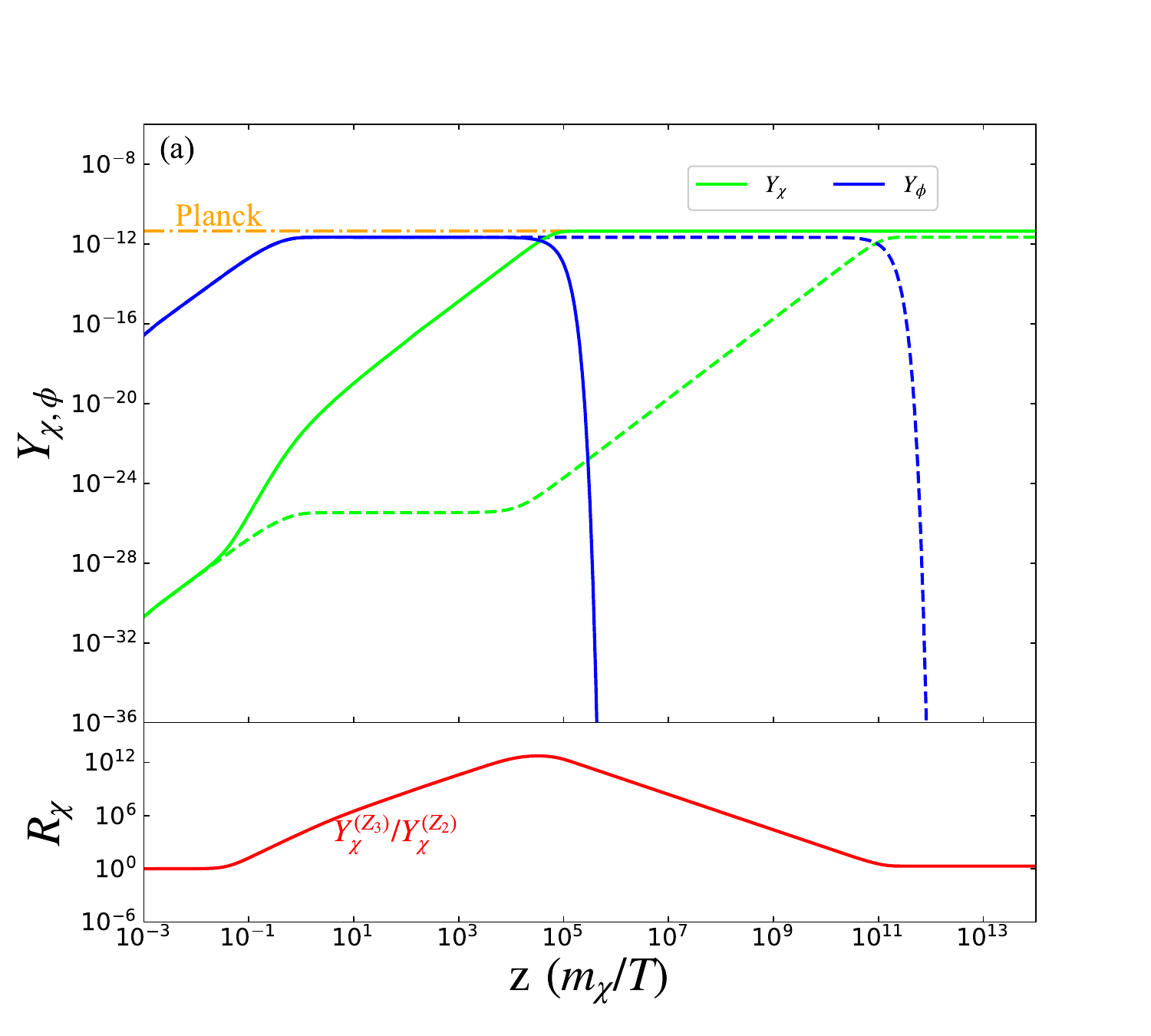}
		\includegraphics[width=0.45\linewidth]{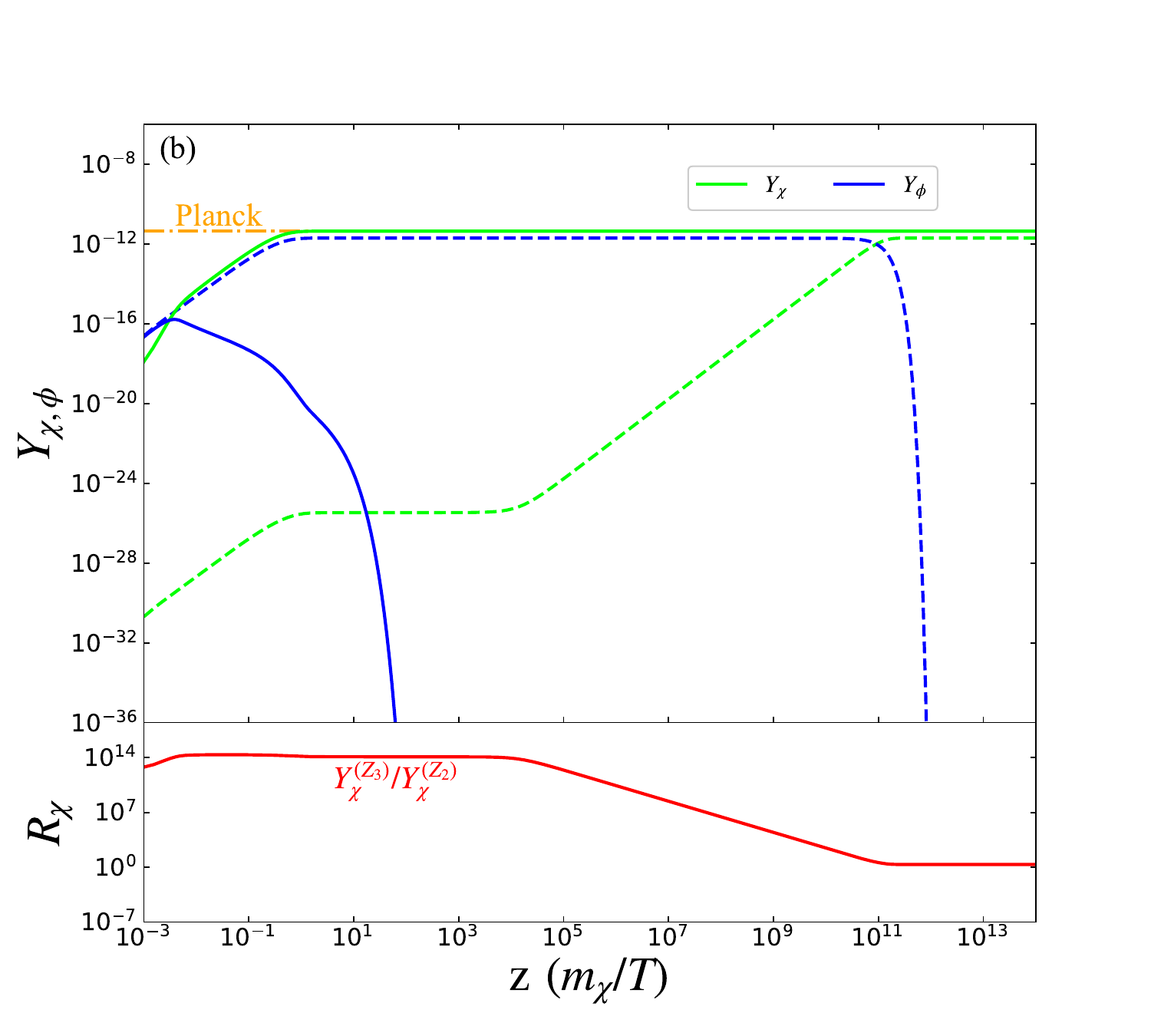}
		\includegraphics[width=0.45\linewidth]{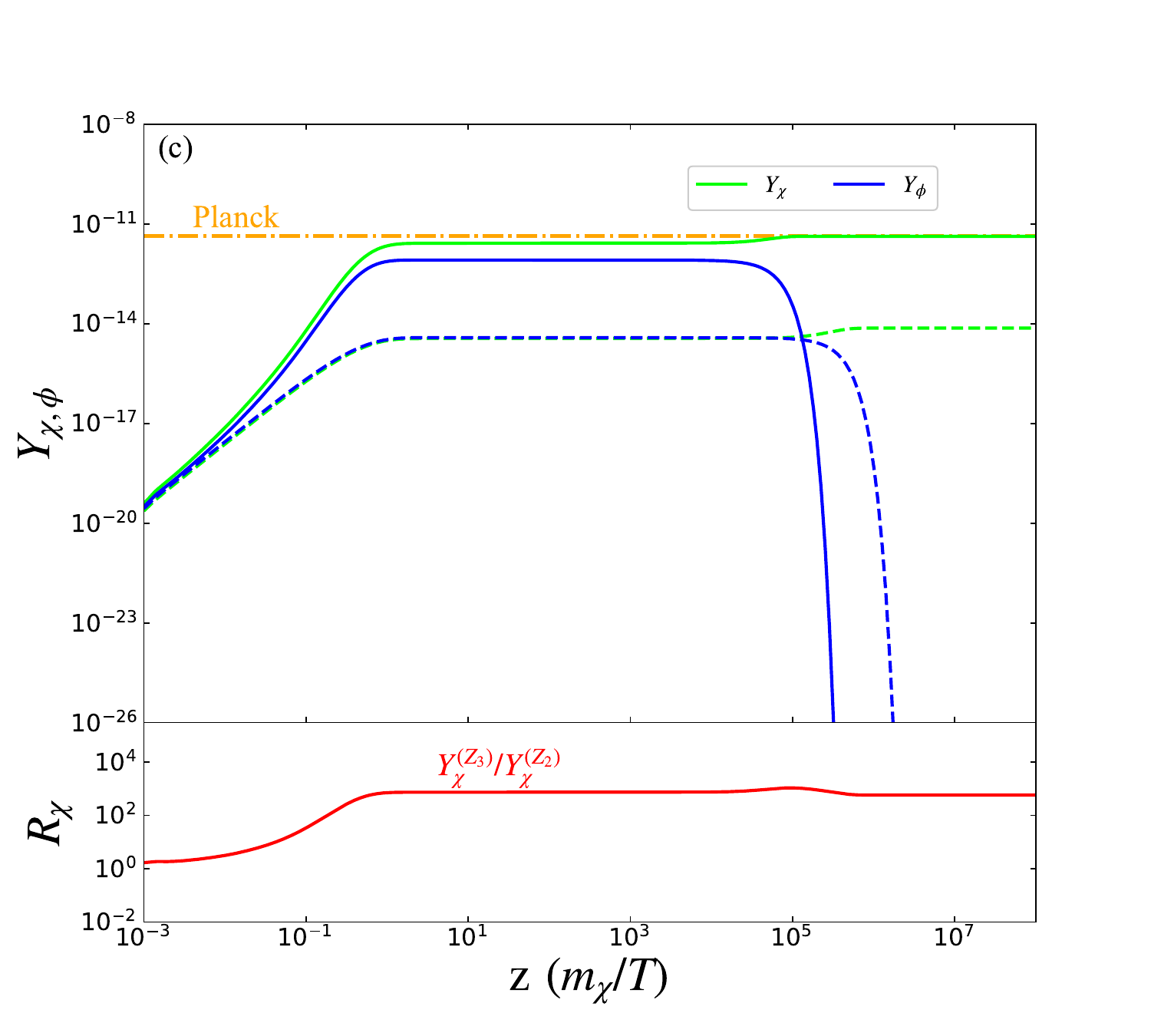}
		\includegraphics[width=0.45\linewidth]{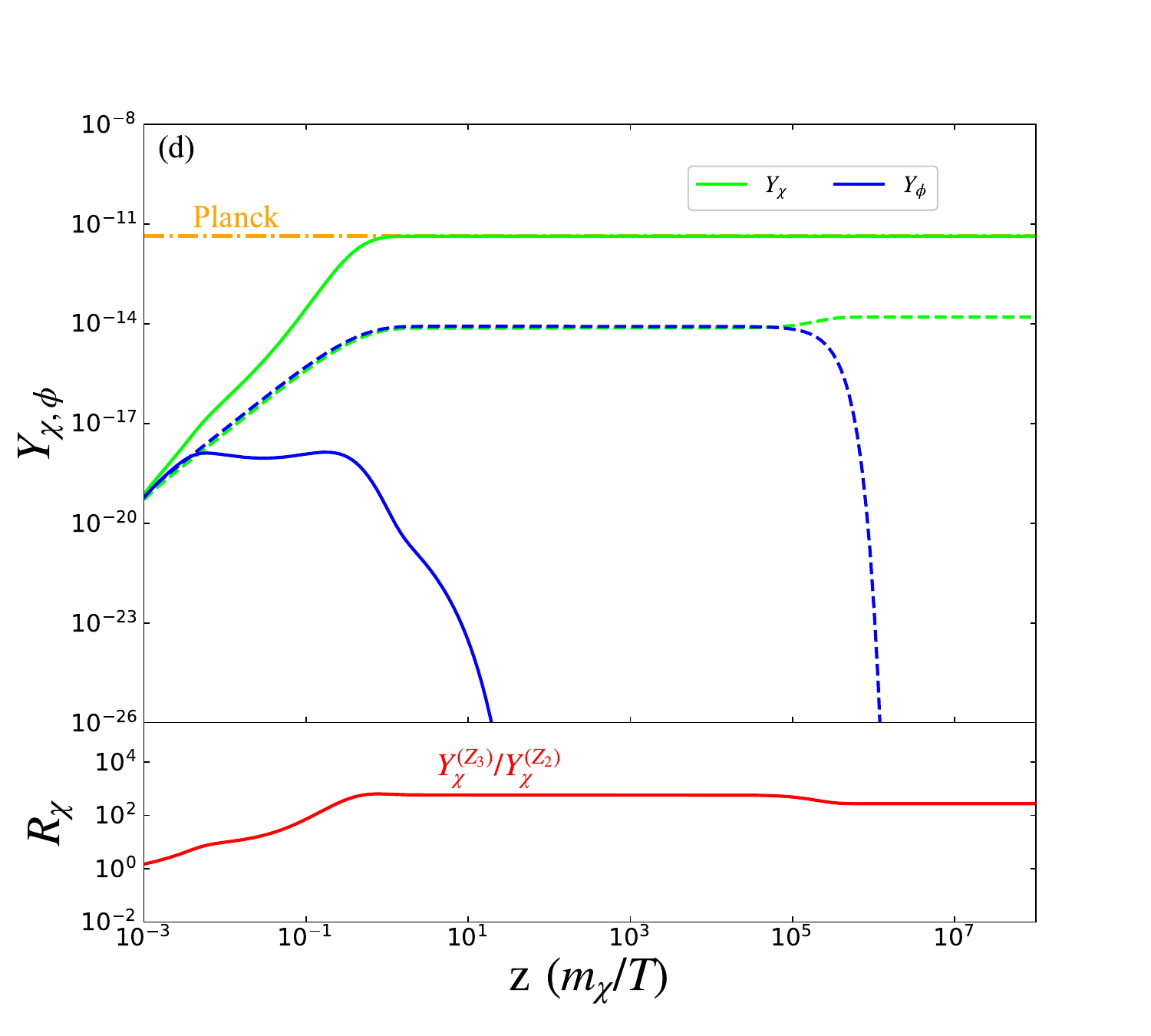}
	\end{center}
	\caption{Same as Figure.~\ref{FIG:fig3}, but for scenario 4.}
	\label{FIG:fig6}
\end{figure}

Scenario 4 has also opened the pair decay $\phi \to \chi\chi$ in contrast with scenario 3. Besides the final decay mode of dark scalar $\phi$, the initial generation channels of the dark sector in scenario 4 are consistent with that in scenario 3.  Table ~\ref{Tab:scenario 4} and Figure~\ref{FIG:fig6} correspond to the selection of parameters and the evolution of dark abundances, respectively.  

In scenario 4 (a), the dark scalar $\phi$ is produced via the Higgs portal $\SM\to\phi\phi$ process. Productions from $2\to2$ scattering processes are quite inefficient, and the DM $\chi$ is generated by the fast pair decay $\phi\to\chi\chi$ under the $Z_3$ symmetry. Compared with scenario 3 (a), a slightly smaller $\lambda_{H\phi}$ is enough to realize the correct DM relic abundance, which is also due to the pair decay. This decay can lead to the ratio $R_\chi$ increasing to $\mathcal{O}(10^{12})$, and then decreasing to two finally.

In scenario 4 (b), both the conversion process $\phi\phi\to\chi\chi$ and decay $\phi\to\chi\chi$ are greatly enhanced. Same as in scenario 2 (b), these two processes lead to more efficient production of DM than scenario 4 (a), so a smaller $\lambda_{H\phi}$ in this scenario is enough to produce correct DM abundance. The ratio $R_\chi$ quickly reaches the maximum value of $\sim10^{14}$, then gradually decreases to 2.2.

In scenario 4 (c), the dark sector abundances are firstly generated by the $2\to2$ scattering processes with typical cross section $\langle \sigma v\rangle_{NN \to \phi\phi}\simeq 2.2\times10^{-49} ~\rm{cm^{3}/s}$, $\langle \sigma v\rangle_{NN \to \chi\chi}\simeq 7.4\times10^{-50} ~\rm{cm^{3}/s}$ and  $\langle \sigma v\rangle_{\chi\phi \to h\nu}\simeq 1.0\times10^{-47} ~\rm{cm^{3}/s}$ for the benchmark point. Then the relatively large semi-production processes $N\chi\to\phi\phi$ and $N\phi\to \phi \chi$ exponentially enhance the dark sector abundances. The ratio $R_\chi$ exponentially increases to  $7.3\times10^{2}$, and is further enlarged by the pair decay $\phi\to\chi\chi$. Finally $R_\chi$ decreases to $5.8\times10^{2}$ due to the delayed contribution of $\phi\to\chi\nu$ under the $Z_2$ symmetry.

In scenario 4 (d), the large pair decay width $\Gamma_{\phi\to\chi\chi}$ makes the dark scalar $\phi$ quite short-lived. The produced dark scalar rapidly decays into the DM pair, rather than taking part in the semi-production processes $N\chi\to\phi\phi$ and $N\phi\to \phi \chi$, which clearly weakens the exponential enhancement effect. Therefore, a larger $y_N$ is required to produce the observed DM abundance compared with scenario 4 (c).  The ratio $R_\chi$ exponentially increases to $5.8\times10^{2}$, then decreases to $2.9\times10^{2}$ finally.

Similar to scenario 2, the pair decay $\phi\to\chi\chi$ is more efficient in producing DM abundance in the $Z_3$ symmetric model even when the dark scalar is generated through the Higgs portal $\SM\to\phi\phi$. Exponential enhancement by the semi-production processes $N\chi\to\phi\phi$ and $N\phi\to \phi \chi$ are also possible in this scenario. However, the rapid pair decay $\phi\to\chi\chi$ may weaken the enhancement effect. 

\section{Phenomenology} \label{SEC:CP}

The sterile neutrino portal FIMP DM model has rich phenomenology \cite{Liu:2022cct}. Despite the DM $\chi$ being hard to detect, both the sterile neutrino $N$ and dark scalar $\phi$ lead to observable signatures. The sterile neutrino $N$ can be directly produced at colliders \cite{Abdullahi:2022jlv}. Meanwhile, the neutrino from delayed decay $\phi\to\chi\nu$ affects the Cosmic Microwave Background (CMB), the energetic neutrino spectrum  and the effective number of relativistic neutrino species \cite{Liu:2022cct}.  

\begin{figure} 
	\begin{center}
		\includegraphics[width=0.45\linewidth]{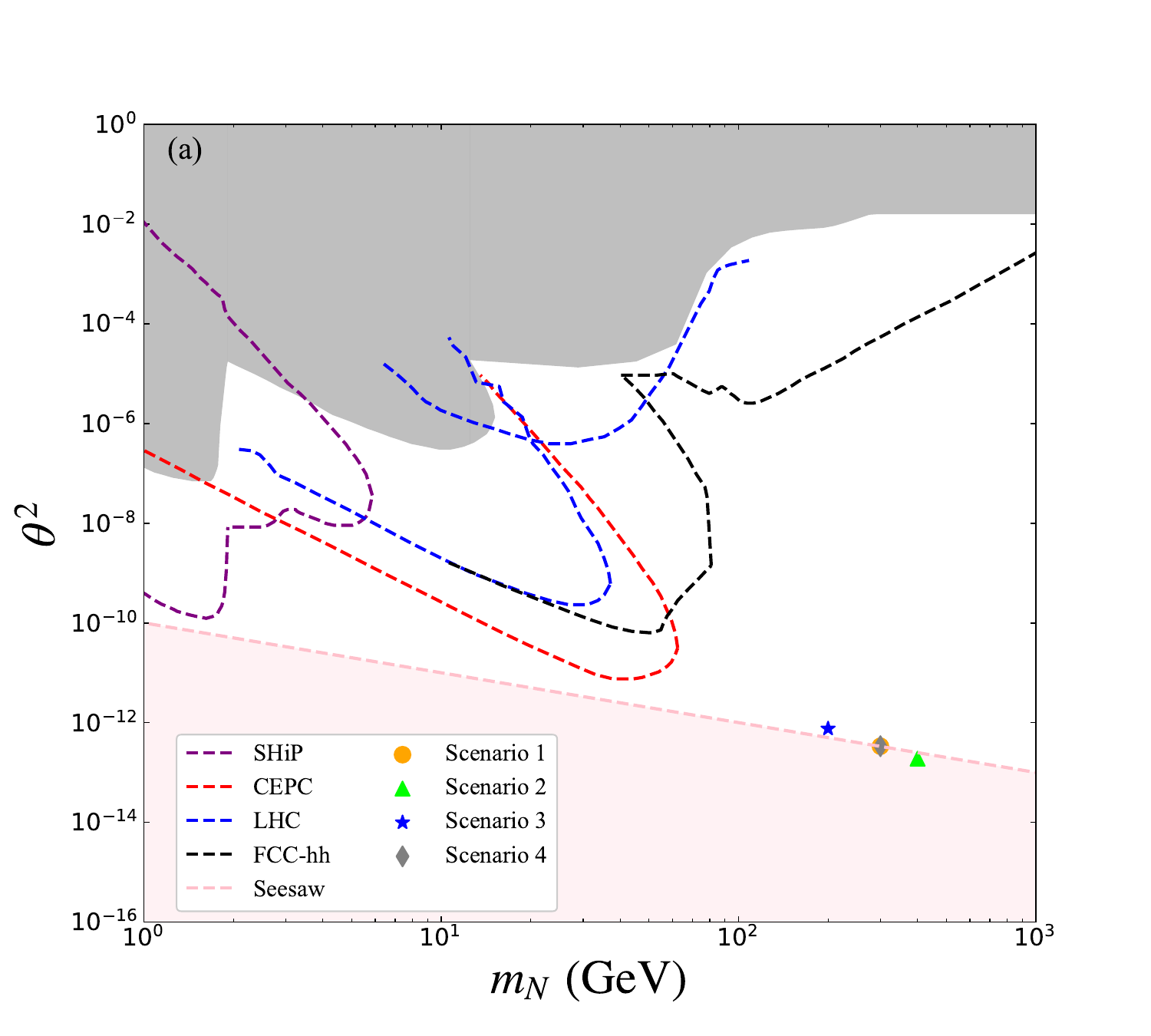}
		\includegraphics[width=0.45\linewidth]{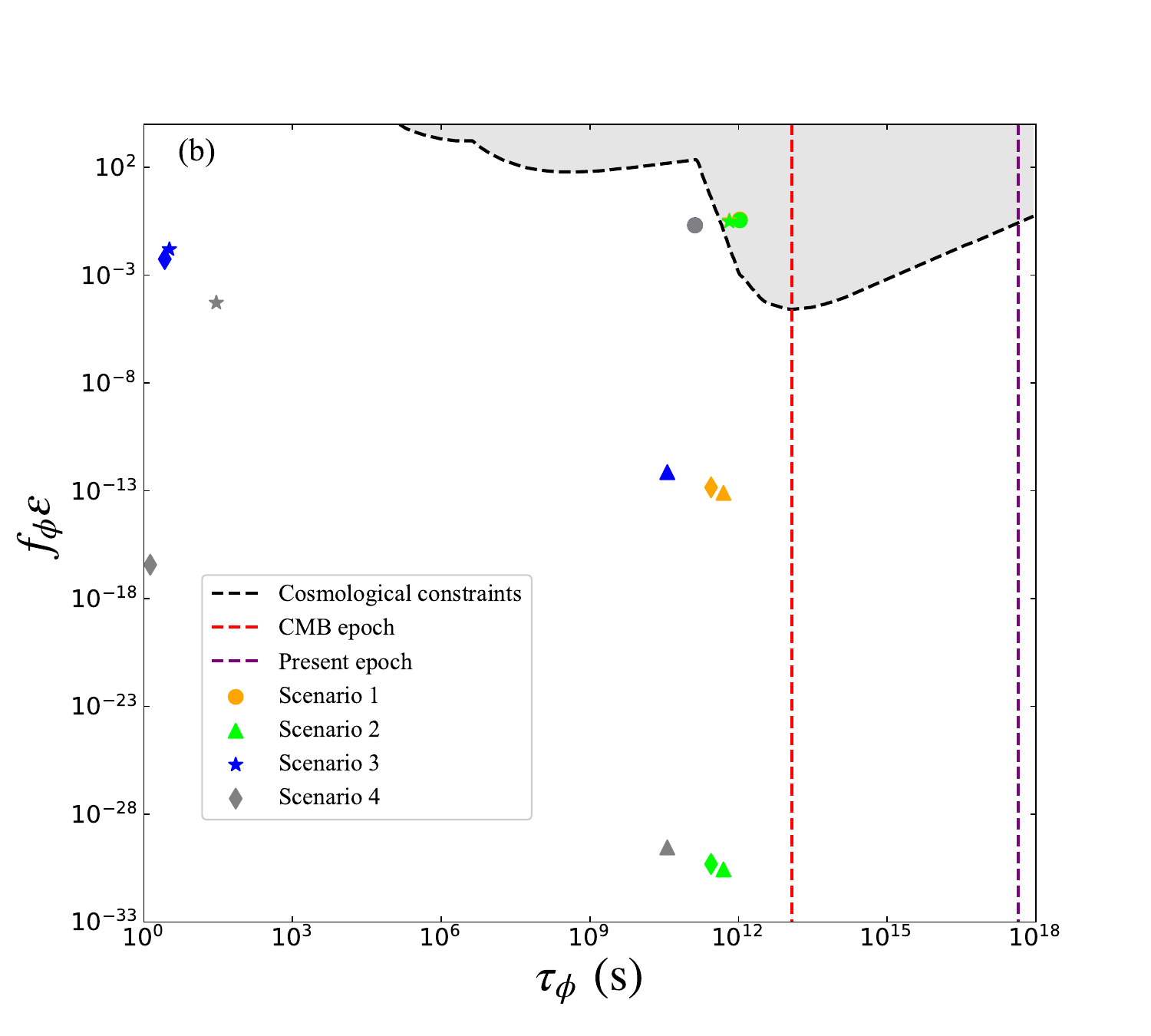}
	\end{center}
	\caption{Status and future prospect of sterile neutrino $N$ (left). The gray areas have been excluded by current experiments~\cite{Blondel:2022qqo}. The purple, red, blue and black dashed lines are the future limits from SHiP~\cite{SHiP:2018xqw, Gorbunov:2020rjx}, CEPC~\cite{CEPCStudyGroup:2018ghi}, LHC~\cite{Pascoli:2018heg, Izaguirre:2015pga} and FCC-hh~\cite{Antusch:2016ejd}, respectively. The pink line indicates the seesaw predicted limit. Cosmological constraints of dark scalar $\phi$ (right). In the right panel, the black dotted line represents the cosmological constraint discussed in \cite{Hambye:2021moy} with $m_\phi = 100~\GeV$, the red and purple dotted lines represent the two epochs of CMB and present, respectively. The circle, triangle, star, and diamond represent scenarios 1 to 4. Meanwhile, the orange, green, blue and gray samples represent the four cases (a) to (d) for each scenario.}
	\label{FIG:fig7}
\end{figure}

The collider signatures of sterile neutrino $N$ will be analyzed briefly. The electroweak scale $N$ can be produced at LHC via the process $pp\to W\to \ell^\pm N$. The cross section of this process is determined by the mixing angle $\theta$. Lepton number violation signature arises from the decay $N\to\ell^\pm W^\mp\to \ell^\pm q_1 \bar{q}_2$ \cite{Cai:2017mow}. When $m_N<m_W$, the three-body decay via off-shell $W/Z$ is the dominant channel, which leads to the displaced vertex signature \cite{Drewes:2019fou}. In Figure \ref{FIG:fig7} (a), we summarize the status and future prospect of $N$. By searching for the displaced vertex signature, a quite large part of the parameter space with $m_N<m_W$ can be covered in the future. For our benchmark scenarios, $y_\nu=10^{-6}$ corresponds to $\theta^2\sim10^{-12}$ with a natural seesaw relation, which is clearly beyond the scope of future sensitivity. Of course,  a larger mixing angle is possible by tuning the parametrization parameters \cite{Barman:2022scg}. 

Then we will focus on the cosmological constraints on $\phi \to \chi \nu$ in different scenarios under the $Z_3$ symmetry. The secondary particles emitted by the neutrino from delayed decay $\phi\to\chi\nu$ have a great impact on the CMB anisotropies and spectral distortions. In Figure \ref{FIG:fig7} (b), we show the corresponding cosmological constraints, where the fractional abundance $f_\phi= \Omega_\phi/\Omega_{\rm DM}$, $\varepsilon=(m^2_{\phi}-m^2_{\chi})/2m^2_{\phi}$ denotes the fraction of the energy of $\phi$ that has been transferred to neutrinos~\cite{Blackadder:2014wpa}. According to Table ~\ref{Tab:scenario 1} and Figure \ref{FIG:fig3}, scenario 1 (a) and 1 (b) predict the same results, thus are overlapped in Figure \ref{FIG:fig7} (b). The same is true for scenario 1 (c), 1 (d) and scenario 3 (a), 3 (b).

In scenario 1, the typical lifetime of dark scalar $\tau_{\phi}$ is about $10^{11}\sim10^{12}$ s with the tiny coupling $y_N\sim10^{-12}$. While scenarios 1 (a) and 1 (b) are excluded by CMB constraint, scenarios 1 (c) and 1 (d) are still marginally allowed. The DM relic density from $N\to\phi\chi$ has limited the coupling $y_N\lesssim10^{-12}$, so the simplest way to improve scenario 1 is increasing $y_\nu$ to about $\mathcal{O}(10^{-5})$.
In scenario 3, we have $\tau_{\phi}\sim10^{11}$ s for case (a) and (b), meanwhile $\tau_{\phi}\sim10^1$ s for case (c) and (d), respectively. The former two cases are also excluded. Different from scenario 1, increasing $y_N$ for scenarios 3 (a) and (b) is also viable, because the DM relic density from scattering processes only requires $y_N\lesssim10^{-7}$. As for scenarios 2 and 4, the branching ratio of $\phi \to \chi \nu$ is much smaller than that of $\phi \to \chi\chi$, which results in only a tiny part of $\phi$ decaying into neutrinos. Therefore, scenarios 2 and 4 can easily satisfy the cosmological constraints. In the following discussion, scenario 1 and scenario 3 are considered preferentially as $\phi \to\chi\nu$ is the only decay channel. 

\begin{figure}
	\begin{center}
		\includegraphics[width=0.45\linewidth]{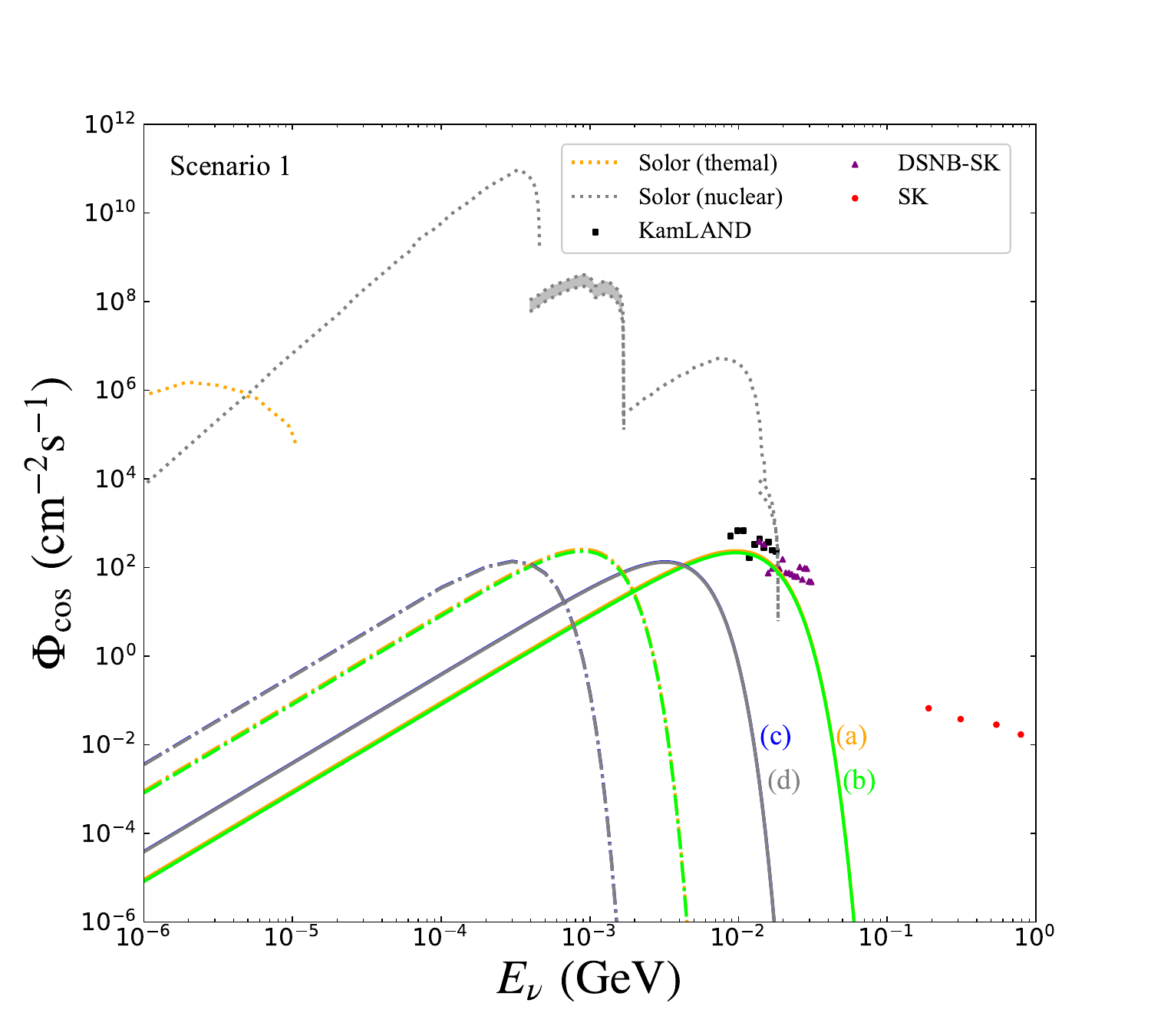}
		\includegraphics[width=0.45\linewidth]{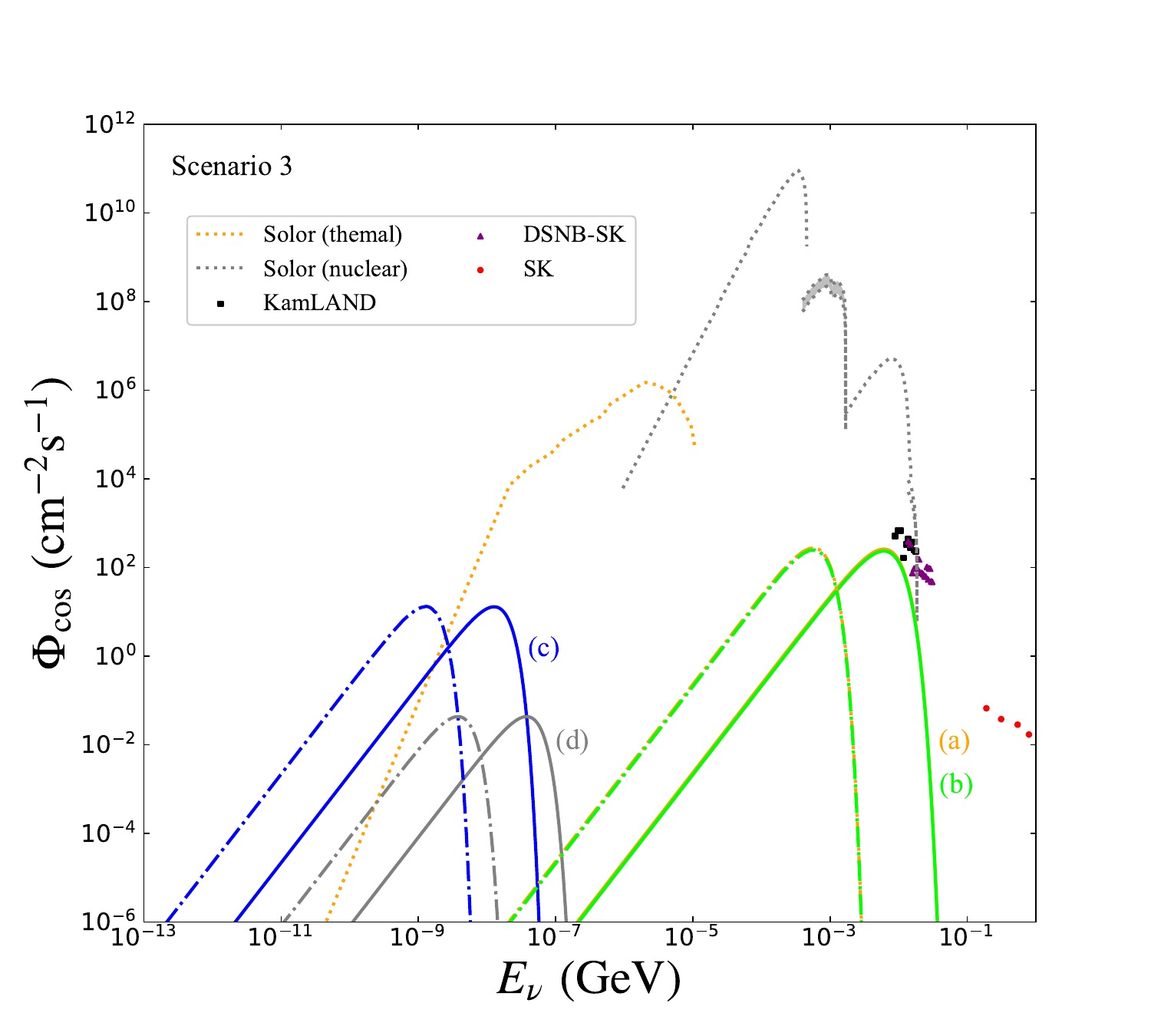}
	\end{center}
	\caption{The predicted neutrino fluxes at present for scenarios 1 and 3. The yellow and gray dotted lines are the thermal and nuclear solar neutrino flux \cite{Vitagliano:2019yzm}, The black squares and purple triangles represent the diffuse supernova neutrino background (DSNB) flux measured at the KamLAND \cite{KamLAND:2011bnd} and SK \cite{Super-Kamiokande:2013ufi}, respectively. The red points are the atmospheric neutrino data from SK \cite{Super-Kamiokande:2015qek}.  The orange, green, blue, and gray solid lines correspond to case (a), (b), (c) and (d) for each scenario, while solid and dot-dashed lines correspond to $y_\nu=10^{-6}$ and $y_\nu=10^{-5}$ respectively.}
	\label{FIG:fig8}
\end{figure}

The energetic neutrinos generated by the delayed decay of $\phi$ will be captured by current neutrino experiments. The neutrino flux at present is calculated as \cite{Bandyopadhyay:2020qpn}
\begin{equation}
	\Phi_{\rm cos}\equiv E_\nu \frac{d\varphi}{dE_\nu}=\left(\frac{n_{\phi}}{\tau_{\phi}}\right)\left(\frac{e^{-t(x)/\tau_{\phi}}}{H(x)}\right)\theta^{'}(x),
\end{equation}
where $E_\nu$ is the observed neutrino energy, $d\varphi/dE_\nu$ is the predicted neutrino flux, $n_{\phi}$ is the number density of $\phi$ if it is stable, $\theta^{'}(x)$ is the Heaviside theta function. The cosmic time $t(x)$ at red-shift $1+x$ and the Hubble parameter $H(x)$ in the standard cosmology are given by
\begin{eqnarray}\label{nf2}
	t(x)&\approx& \frac{4}{3H_0}\left(\frac{\Omega_{\rm r}^{3/2}}{\Omega_{\rm m}^{2}}\right)\left(1-\left(1-\frac{\Omega_{\rm m}}{2(1+x)\Omega_{\rm r}}\right)\sqrt{1+\frac{\Omega_{\rm m}}{(1+x)\Omega_{\rm r}}}\right),  \\
	H(x)&=&H_0\sqrt{\Omega_\Lambda+(1+x)^3\Omega_{\rm m}+(1+x)^4\Omega_{\rm r}},
\end{eqnarray}
where $x=E_0/E_\nu-1$ with initial energy $E_0=(m_{\phi}^2-m_{\chi}^2)/2m_{\phi}$, the Hubble constant $H_0=100h~\rm{km/s/Mpc}$ with $h=0.6727$ \cite{Planck:2018vyg}. The dark energy, matter and radiation fractions are $\Omega_\Lambda=0.6846, \Omega_{\rm m}=0.315$ and $\Omega_{\rm r}=9.265\times10^{-5}$, respectively.

The neutrino fluxes generated in scenarios 1 and 3  are shown in Figure~\ref{FIG:fig8}.  The predicted neutrino fluxes with $y_\nu=10^{-6}$ for both scenarios are allowed by current observation. However, such small $y_\nu$ may not be favored by CMB constraint, so we also show the results of $y_\nu=10^{-5}$. A larger $y_\nu$ leads to the dark scalar decaying earlier, resulting in less energetic neutrino flux at present.

The neutrinos generated from $\phi \to \chi \nu$ also increase
the effective number of relativistic neutrino species $N_{\rm eff}$, which can be written as
\begin{equation}\label{Neff}
	N_{\rm eff}=\frac{7}{8}\left(\frac{11}{4}\right)^{4/3}\left(\frac{\rho_\nu}{\rho_\gamma}\right)=3\left(\frac{11}{4}\right)^{4/3}\left(\frac{T_\nu}{T_\gamma}\right)^4,
\end{equation}
where $\rho_\nu$ and $\rho_\gamma$ represent the energy densities of light neutrinos and photons respectively, $T_\nu$ and $T_\gamma$ are their corresponding temperatures.
By modifying the evolution equations of $T_\nu$ and $T_\gamma$ in SM~\cite{EscuderoAbenza:2020cmq, Escudero:2018mvt}, the corresponding equations that conform to our model are
\begin{eqnarray}\label{Tnu}
	\frac{dT_{\gamma}}{dt}&=&-\frac{4H\rho_{\gamma}+3H(\rho_e+p_e)+\frac{\delta\rho_{\nu_e}}{\delta t}+2\frac{\delta\rho_{\nu_\mu}}{\delta t}-\varepsilon\xi_{\rm EM}\frac{\rho_{\phi}}{\tau_{\phi}}}{\frac{\partial\rho_{\gamma}}{\partial T_{\gamma}}+\frac{\partial\rho_{e}}{\partial T_{\gamma}}},\\
	\frac{dT_{\nu}}{dt}&=&-H T_{\nu}+\frac{\frac{\delta\rho_{\nu_e}}{\delta t}+2\frac{\delta\rho_{\nu_\mu}}{\delta t}+\varepsilon(1-\xi_{\rm EM})\frac{\rho_{\phi}}{\tau_{\phi}}}{3\frac{\partial\rho_{\nu}}{\partial T_{\nu}}}.
\end{eqnarray}
where $\rho_{\gamma,e,\nu}$ denote the energy densities of $\gamma$, $e$ and $\nu$. $\rho_{\phi}$ expresses the energy density of $\phi$ provided it is stable. $p_e$ is the pressure density of $e$. $\xi_{\rm EM}$ represents the energy fraction that the neutrinos inject into electromagnetic plasma, which is assumed to be zero for the selection of $m_\phi$ in this work \cite{Hambye:2021moy}. The neutrino-electron energy density transfer rate $\delta\rho_{\nu}/\delta t$ is taken from Refs.~\cite{Escudero:2018mvt, EscuderoAbenza:2020cmq}. In addition, we don't distinguish the flavor of neutrinos here.

\begin{figure}
	\begin{center}
		\includegraphics[width=0.45\linewidth]{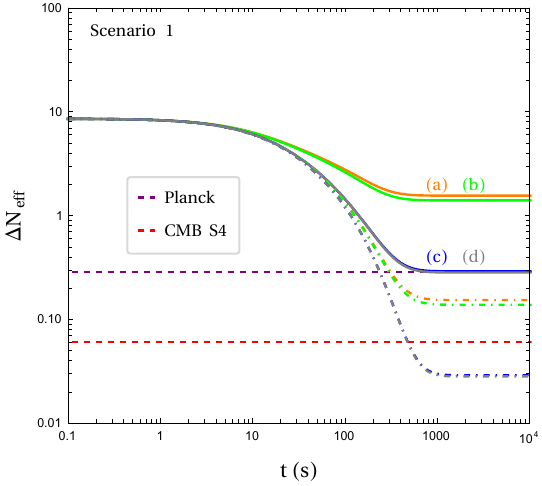}
		\includegraphics[width=0.45\linewidth]{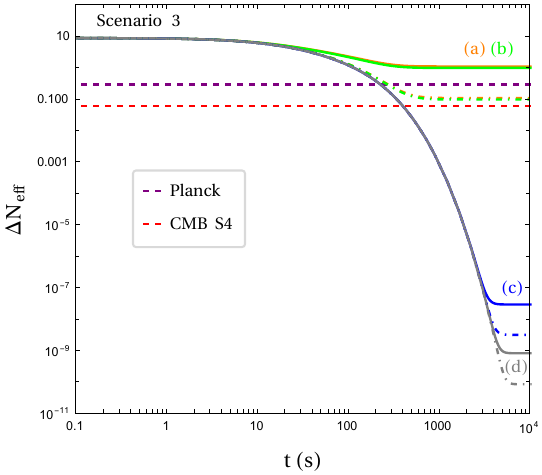}
	\end{center}
	\caption{The evolution of $\Delta N_{\rm eff}$ for scenarios 1 and 3. The calculations are started at $T_\gamma=T_\nu=10$ MeV with the corresponding initial time  $t_0=\frac{1}{2H}|_{T=10 ~\rm MeV}$. The purple and red dashed lines represent the constraints of $\Delta N_{\rm eff}$ from current Planck~\cite{Planck:2018vyg} and future CMB S4~\cite{Abazajian:2019eic}, respectively. The orange, green, blue, and gray solid lines correspond to case (a), (b), (c) and (d) for each scenario, while solid and dot-dashed lines correspond to $y_\nu=10^{-6}$ and $y_\nu=10^{-5}$ respectively.}
	\label{FIG:fig9}
\end{figure}

The evolution of $\Delta N_{\rm eff}$ for scenario 1 and 3 are shown in Figure~\ref{FIG:fig9}, here $\Delta N_{\rm eff}\equiv N_{\rm eff}-N^{\rm SM}_{\rm eff}$ with $N^{\rm SM}_{\rm eff}=3.045$ \cite{Mangano:2005cc, Grohs:2015tfy, deSalas:2016ztq}. For scenario 1, results with $y_\nu=10^{-6}$ are not favored by current Planck observation. When $y_\nu=10^{-5}$, scenarios 1 (a) and 1 (b) predict $\Delta N_\text{eff}\simeq0.14$, which is allowed by current Planck limit and can be further confirmed by future CMB S4. Scenario 1 (c) and 1 (d) with $y_\nu=10^{-5}$ predict $\Delta N_\text{eff}\simeq0.03$, which is beyond the future sensitivity. For scenario 3, cases (a) and (b) with $y_\nu=10^{-6}$ are excluded by Planck, but are still allowed with $y_\nu=10^{-5}$. Case (c) and (d) predict vanishing small $\Delta N_\text{eff}$ even with $y_\nu=10^{-6}$, thus are also hard to detect.

\section {Discussion and  Conclusion } \label{SEC:DC}

The feeble sterile neutrino portal DM with $Z_3$ symmetry is studied in this paper. Besides the sterile neutrino $N$, a dark sector with one fermion singlet $\chi$ and one scalar singlet $\phi$ is also introduced. The dark sector $\phi$ and $\chi$ are charged under a $Z_3$ symmetry. In addition to the well-studied sterile neutrino portal Yukawa coupling $y_N\phi\bar{\chi}N$ and Higgs portal coupling $\lambda_{H\phi}(H^\dag H)(\phi^\dag\phi)$ in the $Z_2$ symmetric model, the $Z_3$ symmetry further allows the dark sector Yukawa interaction $y_\chi \phi \bar{\chi^c} \chi$ and dark scalar self-interaction $\mu\phi^3/2$. Provided the fermion singlet $\chi$ as the FIMP DM candidate, the latter two terms could generate new production channels for DM in the $Z_3$ symmetric model.

Because various production channels depend on the mass spectrum, we consider four specific scenarios to illustrate the evolution of the dark sector. We find that the dominant production and decay mode of dark scalar $\phi$ has a great effect on the evolution of DM. When the delayed decay $\phi\to\chi\nu$ is the only decay mode of $\phi$, the dark scalar generated from the Higgs portal process $\SM\to\phi\phi$ (as in scenario 1 (a), 1 (b), 3 (a), 3 (b)) or from direct decay $N\to\phi\chi$ (as in scenario 1 (c), 1 (d)) will lead to the same final DM abundance for both $Z_2$ and $Z_3$ symmetry, although the conversion process $\phi\phi\to\chi\chi$ could alert the evolution of DM in the $Z_3$ symmetric model. For natural seesaw required Yukawa coupling $y_\nu=10^{-6}$, these involved scenarios with the only delayed decay $\phi\to\chi\nu$ mode are already excluded by cosmological constraints. We show that increasing $y_\nu=10^{-5}$ is sufficient to satisfy all current constraints.

When the pair decay $\phi\to\chi\chi$ is kinematically allowed, it becomes the dominant decay mode of dark scalar, since the delayed decay $\phi\to\chi\nu$ is heavily suppressed by the tiny mixing angle $\theta\sim10^{-6}$ in our analysis. This pair decay $\phi\to\chi\chi$  only appears in the $Z_3$ symmetric model, thus definitely leads to a difference between the two kinds of symmetric models. When the dark scalar is dominantly produced from the Higgs portal process $\SM\to\phi\phi$ (as in scenario 2 (a), 2 (b), 4 (a), 4 (b)), the final DM abundance in the $Z_3$ symmetry is about twice as large as it in the $Z_2$ symmetry. Meanwhile, if the dark scalar is generated from the direct decay $N\to\phi\chi$ (as in scenario 2 (c), 2 (d)), the DM relic abundance ratio of $Z_3$ symmetry to $Z_2$ symmetry is three to two. In short, the pair decay is more efficient in producing DM. With a suppressed branching ratio of $\phi\to\chi\nu$, these scenarios are easily to avoid the cosmological constraints.

The most interesting scenario is when the dark sector is primarily generated by the scattering processes as $NN\to\chi\chi,NN\to\phi\phi,h\nu\to\chi\phi$ (as in scenario 3 (c), 3 (d), 4 (c), 4 (d)). Then the semi-production process $N \chi\to\phi\phi, N\phi\to\phi\chi,N\chi\to\chi\chi$ could lead to the exponential growth of dark sector abundances in the $Z_3$ symmetric model. Compared with the $Z_2$ symmetric model, the final DM abundance of such scenarios could be enhanced by two to three orders of magnitudes. Our benchmark points also indicate that the generation of DM $\chi$ is much more efficient than the dark scalar, which results in a tiny fractional abundance $f_\phi$. Meanwhile, the relatively large Yukawa coupling $y_N\sim\mathcal{O}(10^{-7})$ significantly reduces the lifetime of the dark scalar $\phi$. These two aspects make such scenarios hard to probe via the cosmological observables, even when $\phi\to\chi\nu$ is the only decay mode.

\section*{Acknowledgments}
This work is supported by the National Natural Science Foundation of China under Grant No. 11975011, 11805081 and  11635009, Natural Science Foundation of Shandong Province under Grant No. ZR2019QA021 and ZR2022MA056, the Open Project of Guangxi Key Laboratory of Nuclear Physics and Nuclear Technology under Grant No. NLK2021-07.
%%%%%%%%%%%%%%%%%%%%%%%%%%%%%

\end{document}